\documentclass[11pt,onecolumn]{aa} 

\usepackage{graphicx}
\usepackage{wrapfig}
\graphicspath{ {./images2050/} }	
\usepackage{geometry}
\usepackage{caption}
\usepackage{subcaption}
\usepackage{enumitem}
\usepackage{natbib}
\usepackage{amssymb}
\usepackage{amsmath}
\usepackage{amsfonts}
\usepackage{url}
\usepackage{longtable}
\usepackage{hyperref}
\usepackage{mathptmx}
\usepackage{alltt}
\usepackage{listings}
\usepackage{lineno}
\usepackage{txfonts}
\RequirePackage{array}
\usepackage{makeidx}
\usepackage{multirow}
\usepackage[table]{xcolor}
\usepackage{booktabs,xspace}
\usepackage[T1]{fontenc}
\usepackage[font=small,labelfont=bf,tableposition=top]{caption}
\usepackage{xcolor}
\newcommand{\ignore}[2]{\hspace{0in}#2}

\def\apjl{ApJ}%
\def\apjs{ApJS}%
\def\ao{Appl.~Opt.}%
\def\aap{A\&A}%
\DeclareCaptionLabelFormat{andtable}{#1~#2  \&  \tablename~\thetable}
\renewcommand{\vec}[1]{\ensuremath{\mathchoice{\mbox{\boldmath$\displaystyle#1$}}
		{\mbox{\boldmath$\textstyle#1$}}
		{\mbox{\boldmath$\scriptstyle#1$}}
		{\mbox{\boldmath$\scriptscriptstyle#1$}}}}

\begin{document}
\raggedbottom

\begin{center}
	{\Huge\bf{
			{ Voyage 2050 White Paper \\ \vspace{2cm} All-Sky Visible and Near Infrared Space Astrometry\\}
	}}
	{\LARGE{
			{\vspace{3cm} Principal Applicant: \\ \vspace{1cm} David Hobbs \\ \vspace{1cm} Lund Observatory \\ Box 43 22100 \\ Lund \\ Sweden \\ \vspace{0.5cm}  Email: david@astro.lu.se \\ \vspace{0.5cm} Tel.: +46-46-22\,21573}
	}}
\end{center}
\thispagestyle{empty}
\pagestyle{plain}
\pagenumbering{arabic}

\newpage
\setcounter{page}{1}
\section{Executive summary}\label{sec:summary}

The era of all-sky space astrometry began with the Hipparcos mission in 1989 and provided the first very accurate catalogue of
apparent magnitudes, positions, parallaxes and proper motions of 120 thousand bright stars at the milliarcsec (or milliarcsec per year) accuracy
level. Hipparcos has now been superseded by the results of the Gaia mission, the second Gaia data release contained
astrometric data for almost 1.7 billion sources with tens of microarcsec (or microarcsec per year) accuracy in a vast volume of the Milky Way and
future data releases will further improve on this. Gaia has just completed its nominal 5 year mission (July
2019), but is expected to continue in operations for an extended period of an additional 5 years through to mid 2024. Its final
catalogue to be released $\sim$~2027 will  provide astrometry for  $\sim$~2 billion sources, with astrometric precisions reaching
10 microarcsec. Why is accurate astrometry so important? The answer is that it provides fundamental data which
underpins much of modern observational astronomy as will be detailed in this white paper. All-sky visible and Near-InfraRed (NIR)
astrometry with a wavelength cutoff in the K-band is not just focused on a single or small number of key science cases. Instead,
it is extremely broad, answering key science questions in nearly every branch of astronomy while also providing a dense and
accurate visible-NIR reference frame needed for future astronomy facilities.

\begin{figure}[tbh]
	\begin{center}
		\includegraphics[scale=0.21,trim={4cm 0cm 4cm 0cm},clip]{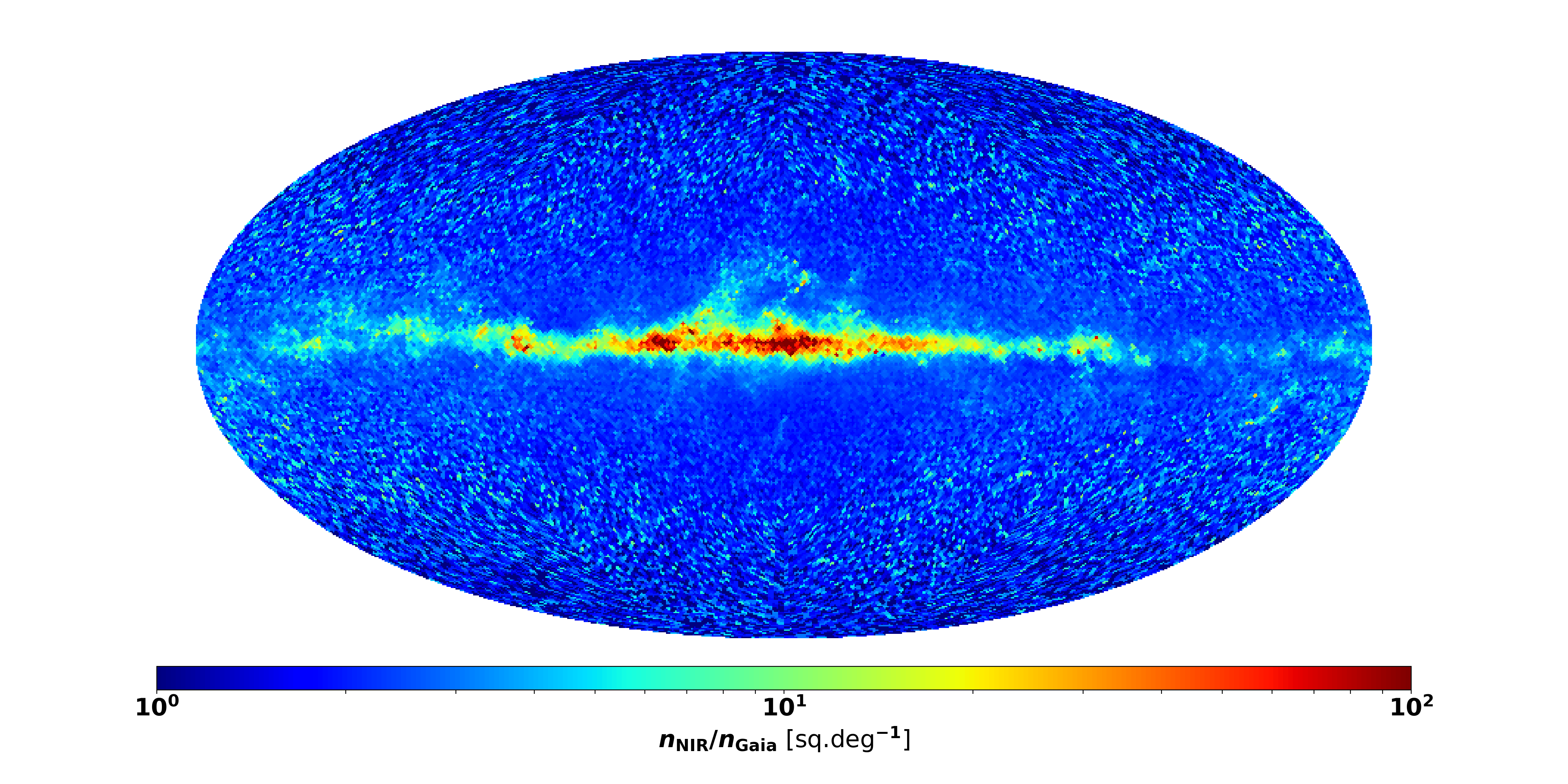}
		\includegraphics[scale=0.21,trim={4cm 0cm 4cm 0cm},clip]{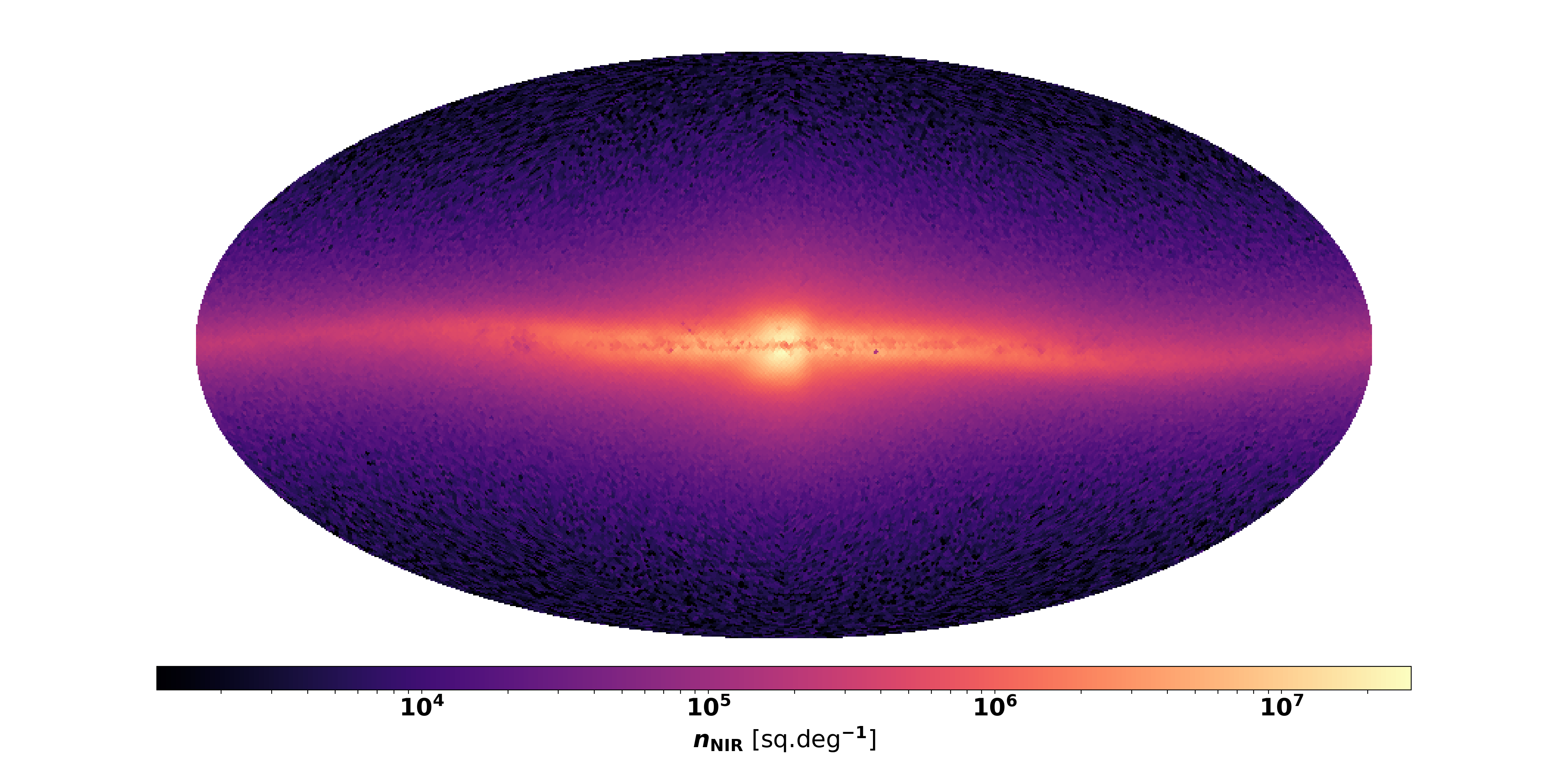}
		\setlength{\belowcaptionskip}{-10pt}
		\caption[]{\em{All-sky projection in Galactic coordinates of the star count ratio per square degree between GaiaNIR and Gaia (G-band limit of 20.7th mag giving 1.5 billion Gaia sources). In total 5 times more stars could be observed, especially in the disk where extinction is highest, by GaiaNIR for the H-band limit of 20th mag (left figure) and 6 times more stars could be observed by including the K-band limit of 20th mag. Crowding is not taken into account here and will limit the increase in numbers  in the densest areas. In the right figure we show the corresponding H-band number densities. The underlying Milky Way model (which does not include clusters or external galaxies, e.g. SMC and LMC) is similar to GDR2mock \citep{2018PASP..130g4101R} (using Galaxia \citep{2011ApJ...730....3S} with the extinction map of \cite{2016ApJ...818..130B})
		but only 0.1\% of the stars are sampled explaining the noise in low density regions. A note of caution, the estimation of the star count ratio between Gaia and GaiaNIR is uncertain due to the uncertainty in the extinction model used (older models gave a lower ratio of around 3), mainly towards the centre of the Galaxy. However, one could argue that this uncertainty is a key science case in itself that cannot be resolved by Gaia alone.}}\label{fig:Hband}
	\end{center}
\end{figure}

For almost 2 billion common stars the combination of two all-sky space observatories would provide an astrometric foundation for all
branches of astronomy -- from the solar system and stellar systems, including exoplanet systems, to compact galaxies, quasars, neutron
stars, binaries and dark matter (DM) substructures. The addition of NIR will result in up to 8 billion newly measured stars in some of the
most obscured parts of our Galaxy, and crucially reveal the very heart of the Galactic bulge region (see Figure \ref{fig:Hband} for a note
of caution!).

In this white paper we argue that rather than improving on the accuracy to answer 
specific science questions, a greater overall science return can be achieved by going deeper than Gaia and by expanding the wavelength range to the NIR. 
An obvious question to ask is -- can a more accurate all-sky mission than Gaia be done? Clearly the answer is yes, if we can build a space
telescope with a larger aperture ($D$) but in practice it is very difficult to do this without greatly inflating the cost of the
mission. Gaia was designed very well and only just fitted in the available launchers so improving the telescopes angular
resolution (i.e. minimum angular separation) $R \propto \lambda/D$ at a fixed wavelength $\lambda$ is very costly. Other mission
proposals have tried to avoid this by employing long focal lengths and advanced metrology systems for ultra-accurate narrow field
proposals, like SIM, NEAT and Theia, but these missions were focused on answering important specific science cases and did not aim
to do a broad all-sky astrometric survey. Nevertheless, the metrology systems explored may  find application in improving a future
all-sky mission.

A new all-sky NIR astrometric mission will expand and improve on the science cases of Gaia using basic astrometry.
Key topics are focused on what dark matter is and how is it distributed, how the Milky Way was formed and how has it been impacted by mergers and collisions?
How do stars form and how does stellar feedback affect star formation; what are the properties of
stars, particularly those shrouded in dust, and small solar system bodies; how are they distributed and what is their motion?
%Are there Earth crossing asteroids which could potentially threaten life on Earth; 
How many co-planar systems like ours (with Earth-sized and giant planets) are there and what fraction have planets with long period orbits?
To answer these questions there are three main science challenges for a new all-sky astrometry mission:

\begin{enumerate}
	\item NIR astrometry (and simultaneous photometry) is crucial for penetrating obscured regions and for observing intrinsically red objects 
	when implemented with sufficient accuracy.  
	Peering through the dust of the Milky Way to obtain a dense sampling of the phase space necessary to study the bulge, bar, bar-disk interface 
	and spiral arms. Spiral structure can excite stars to radially migrate and induce disk heating and accurate measurements of the 3-D motion 
	and properties of these obscured stars are needed to trace the dynamical history and evolution of our Galaxy. Preliminary estimates show that 
	a new NIR astrometry mission would observe at least 5 times as many stars as Gaia, assuming the same magnitude range, giving a huge increase in the catalogue size 
	and would dramatically increased phase space sampling of the disk, especially of the innermost regions where co-existing populations require better statistics. 

	\item A new mission could be combined with the older Gaia catalogue (currently $\sim$~1.7 billion sources) with a 20 year interval to give a much 
	longer baseline of 25--35 years, with very accurate proper motions (a factor of 14--20 better in the two components) and improved parallaxes needed to measure 
	% Note the real baseline can be from 25--35 years for two 5yr or two 10yr missions respectively
	larger distances. Dynamical studies in all of the outer halo would give an enormous sample and greatly enhance the resolving of tangential motions in streams 
	and local dwarf galaxies, with a potential accuracy of 2--3 km~s$^{-1}$ for samples out to $\sim$~100~kpc. This will provide great insight into the 
	gravitational potential in the outer reaches of the Milky Way where halo streams are sensitive probes.  Proper motions will also allow the study of 
	massive local group galaxies (for instance Cen-A). At the same time the parallaxes, especially of binaries, will be much 
	improved when astrometric data from two missions are combined and the detection of planets with significantly longer periods than by Gaia alone can 
	be achieved. A significant population of stars with planetary system architectures similar to our Sun's (so long period, massive gas 
	giants, like Jupiter and Saturn, in the outer reaches, shielding Earth type planets in the star's habitable zone) will be discovered.
	
  \item A new mission would allow the slowly degrading accuracy of the Gaia visible reference frame, which will become the
    fundamental Celestial Reference Frame and the basis for all modern astronomical measurements, to be re-initialised back to a
    maximal precision. This degradation is due to errors in its spin and due to small proper motion patterns which
    are not accounted for. The catalogue accuracy itself will decay more rapidly due to errors in the measured proper motions.
    Dense and accurate reference grids are needed for forthcoming giant telescopes but also for smaller
    instruments currently operating or being planned. The extension of this visible reference frame into the NIR is an important
    step given that so many new space and ground based observatories will have infrared sensitive instruments. A new mission would
    provide better accuracy to explore proper motion patterns (e.g. from Sun's Galactic acceleration and gravitation waves), real
    time cosmology and fundamental physics.
\end{enumerate}

In summary, the new mission proposed here will observe many new stars in obscured regions. We estimate at least 5 times as many stars will be observed, giving up to 8 billion new objects. 
NIR opens up a new wavelength range which allows us to probe the dusty obscured regions of the Galactic disk with high-precision astrometry and broad-band high-resolution photometry, while out of the Galactic plane a new mission will go deeper to enhance the halo science cases and provide complementary legacy data to ground based surveys such as LSST. A common astrometric solution for the two missions will give greatly improved proper motions but also improve the parallaxes, for up to 2 billion common stars.
Long term maintenance and expansion of the dense and very accurate celestial reference frame with a new mission is necessary for future precise 
astronomical observations and provides an essential service for the astronomical community.
These features ensure that a new mission is not simply an increment on the previous one but will create an astrometric revolution in itself!

ESA's study of the proposal for GaiaNIR \citep{2016arXiv160907325H} concluded that the overall cost of the mission
would be similar to that of Gaia. However, as the US is the world leader in detector technology we recently proposed
\citep{2019arXiv190408836M, 2019arXiv190705191H} a collaboration with the US on this project, particularly for the detectors,
which would make the mission feasible and allow it to remain within the ESA M-class mission budget provided the US, Japanese, and
Australian partners contribute significantly. 
%Alternatively, collaboration with the Japanese could provide the missing 
%NIR science cases via a dedicated NIR pointed follow-up mission using relative astrometry, if suitable visible-NIR detectors cannot be developed in time -- again achieving 
%the science goals.   
%\newpage

\section{Science Case Background}\label{sec:sciencecase}
The nominal Gaia mission \citep{2016A&A...595A...1G} will provide global astrometry to unprecedented accuracies
($17$--$2$5~$\mu$as (yr$^{-1}$) at $G=15$), in positions, absolute parallaxes and proper motions, with the addition of all-sky
homogeneous multi-colour photometry and spectroscopy. These unique capabilities go well beyond and are complementary to the
science cases being addressed by ground based surveys (such as RAVE, SDSS (photometry, SEGUE and APOGEE), Gaia-ESO, LAMOST, GALAH,
VISTA (VVV), Pan-STARRS, DES, LSST, WEAVE and 4MOST; see section \ref{sec:synergies} on synergies). A space-based mission avoids the limitations caused by the turbulent
atmosphere and the use of Earth rotation parameters and models of nutation and precession. All-sky space-based astrometry leads to
a global solution and provides a rigid sphere for a celestial reference frame that cannot be accurately obtained with any other
method \citep{2018A&A...616A..14G}.  

An obvious technological improvement to the current Gaia mission is to also go into the non-thermal Near-InfraRed (NIR) with a wavelength 
cutoff in the K-band allowing the new mission to probe deeper through the Galactic dust to observe the structure and kinematics of the star forming 
regions in the disk, the spiral arms and the bulge region, to give model independent distances and proper motions in these obscured parts of 
the sky. Additionally, having two 5 or 10-year Gaia-like missions separated by 20 years would give 14--20 times 
better proper motions for a few billion common stars and also improved parallax determinations with new observations.
% by about 40\%, which is crucial for the bulk of distant objects. 
After the publication of the final Gaia catalogue the positions of stars will be accurately known at the chosen reference 
epoch, which will be close to mid 2019. However, this accurate positional information and the accuracy of the link to the VLBI reference 
frame will slowly degrade due to the small uncertainties in the proper motions of the stars. Hence, it is very desirable 
to repeat the measurements of Gaia after about 20 years to maintain the positional accuracy of the stars and the visible 
reference frame. Gaia is already one of the most transformational missions ever as measured in terms of scientific output. 
The rate at which papers appear using Gaia data is soon to outstrip the HST and a new NIR mission will enable a further explosion of science insight.

\cite{2016arXiv160907325H} proposed to the European Space Agency (ESA) a new all-sky NIR astrometry mission, called GaiaNIR.
Such a NIR space observatory is however not possible today: it requires new types of Time Delay Integration (TDI) NIR detectors to scan the 
entire sky and to measure global absolute parallaxes. Although developing TDI-NIR detectors is a significant challenge, the US is well placed 
to advance such technology, and thus to open the doors to what can become the first international collaboration for a global astrometry space 
observatory. In 2017, \href{http://sci.esa.int/future-missions-department/60028-cdf-study-report-gaianir/}{an ESA study} of the GaiaNIR proposal 
already hinted that a US-European collaboration would be the optimal answer to make GaiaNIR science and technology a reality and subsequently 
\cite{2019arXiv190408836M, 2019arXiv190705191H} submitted white papers to the US decadal survey (Astro 2020) outlining the science cases and a possible US-European collaboration.
The Australian National University is also developing NIR astronomical detector technology with TDI capabilities and are very interested in becoming part of this endeavour.
The Japanese are currently working in a different direction with small-JASMINE, which has been recently selected by ISAS/JAXA for their M-3 mission with a 
current scheduled launch in mid-2020s, to do relative (to Gaia) astrometry in the NIR, but only focusing on the small region within $\sim$~100~pc from the 
Galactic centre and relatively bright (Hw$<15$~mag) stars. With their experience from small-JASMINE they are clearly interested in collaborating on the 
new mission outlined here.  In section \ref{sec:mission} we will return to the possible implementation of our mission.

Below we present the science cases for a new all-sky astrometry mission assuming the associated technological challenges can be resolved in reasonable time. 
The accuracy of the new mission should be at least that of Gaia using tried and trusted instrumentation, techniques, and lessons learned 
from Gaia to unveil a wealth of new and more accurate information about our Galaxy. The science cases have been roughly divided into three 
sections 1) NIR science cases; 2) improved proper motion science cases and 3) reference frame science cases. Clearly there is a strong 
overlap between these three areas but they are presented separately for clarity.

\subsection{Measurement concept}\label{sec:sciencecase:measurement}

The measurement concept for obtaining global astrometric measurements only possible from space is well-defined and well 
demonstrated \citep[see, for example,][]{2001A&A...369..339P, 2018A&A...616A...2L}. Indeed, the concept is identical to that of the currently 
flying Gaia mission where the nominal mission lifetime is 5--6 years. However, for Gaia, the community has already proposed to continue 
observations for a total of 10 years if fuel consumption and the hardware on-board continue to operate as expected. Such a mission will 
effectively double the number of measurements giving an accuracy improvement by a factor of $\sqrt{2}$ in the positions and parallaxes but 
a factor  of $2\sqrt{2}$ in the proper motions which also benefit from a doubling of the measurement baseline. However, in addition to just 
doubling the number of measurements, if we also add a gap between missions, a simple calculation shows that the combination of two 5 year missions  
(labelled with subscript N for GaiaNIR and G for Gaia), assuming a positional and a proper motion accuracy of 25~$\mu$as~(yr$^{-1}$)\footnote{This is 
the assumed end of mission (5 years) accuracy but neglects the fact that proper motions are generally better than positions. On the other hand, we do not account 
for parallax improvements which indirectly improve the joint proper motion solutions. Our AGISLab simulations, discussed above, include all of these effects naturally 
and the results are roughly in agreement with these simplistic predictions.}  with a 
separation of 20 years will 
give \citep[see equation 2 in][]{2015AJ....150..141F}:
\begin{align}
&&
\sigma_{\mu_\alpha *} = \frac{\sqrt{\sigma_{\alpha^*_{\rm N}}^2 + \sigma^2_{\alpha^*_{\rm G}}}}{t_{\rm N} - t_{\rm G}}
= \frac{\sqrt{25^2 + 25^2}}{20} \sim 1.77~\mu{\rm as~yr}^{-1} \, ,
&&
\sigma_{\mu_\delta} = \frac{\sqrt{\sigma_{\delta_{\rm N}}^2 + \sigma^2_{\delta_{\rm G}}}}{t_{\rm N} - t_{\rm G}}
= \frac{\sqrt{25^2 + 25^2}}{20} \sim 1.77~\mu{\rm as~yr}^{-1} \, 
\end{align}
which is a factor of 14 better in both proper motion components. If we then assume two 10 year missions one gets
an extra factor of $\sqrt{2}$ improvement in the individual positions from each mission giving an overall improvement by a factor of 20 
for both proper motion components compared to the initial values of 25~$\mu$as~yr$^{-1}$. If a new mission follows we would get these improvements 
in proper motions for a few billion stars. Parallaxes will 
also improve mainly due to the additional measurements \citep{2014arXiv1408.2190H} but also because the proper motions are much 
more accurate. The parallaxes can then be better determined (an indirect improvement) which has already been demonstrated in 
the TGAS solution \citep{2014A&A...571A..85M, 2015A&A...574A.115M} for Gaia where Hipparcos/Tycho-2 data were combined with Gaia's to form 
the first data release. Residual systematic errors will be present in both Gaia and the new mission, but they should be uncorrelated 
and it may be possible to use a joint solution of both missions to partially reduce these errors.
% Matlab check
%  (25/sqrt(1))/(sqrt((25/sqrt(2))^2 + (25/sqrt(2))^2)/20)

AGISLab numerical simulations \citep[See Fig. 6]{2016arXiv160907325H} of two joint missions using periods of 5 and 10 years for each mission segment, respectively, 
show that, indeed, the improvements in the final astrometric errors for proper motion are roughly factors of 14 and 20 better than a 5 year Gaia mission 
alone. Likewise, the improvements in positions and parallax followed the prediction of $\sqrt{2}$ also mentioned above. The numerical simulations show that 
these simplistic estimates give the correct order of improvement for `real' astrometric missions.
This approach brings the advantage of a largely known and already implemented mission concept \citep{2012A&A...538A..78L}, 
which can be improved based on existing experience (particularly by going into the NIR). 
The lessons learned from Gaia will be invaluable and improvements in the data processing and instrument modelling
can be built on already well developed concepts. To accomplish this goal we need to build a new all-sky astrometry mission 
to fly around 2040.

\section{NIR astrometric science cases}\label{sec:nir}

\subsection{The bulge/bar-disk interface and radial migration}\label{sec:nir:diskinterface}
Since Gaia was proposed, it has become clear that the evolution of the Milky Way is far more complex than had been realized. Not only is it
not in equilibrium (e.g. \cite{2018Natur.561..360A}), but its stars move away from their birth places, a process called radial migration.
Originally it was proposed that radial migration could be triggered by bars  (e.g. \cite{1994ApJ...430L.105F})  or spiral arms
\citep{2002MNRAS.336..785S}. Subsequent studies have shown the migration efficiency to vary with time and distance from the Galactic centre,
and to be strongly influenced by the bar and its interaction with spiral arms \citep{2010ApJ...722..112M, 2011A&A...534A..75B} as well as by
minor mergers (e.g. \cite{2012MNRAS.420..913B}). This has important consequences for the chemodynamical evolution of the Galaxy
\citep{2008ApJ...684L..79R, 2009MNRAS.399.1145S, 2013A&A...558A...9M, 2014A&A...572A..92M}. It implies that even the local volume near the
Sun cannot be understood in isolation, without a proper description of the innermost regions  of the Milky Way and its merger history.

An important element of radial migration happens at the interface between the bar and the spiral arms, which occurs in the inner few kpc of
the Galaxy. \citet{2016MNRAS.460L..94G} demonstrated that radial migration can be captured as systematically different stellar motions by
about 10~km~s$^{-1}$ at different sides of the spiral arm. It also needs to be tested at a large range of radii, because it is expected to
be strong at the resonances \citep{2002MNRAS.336..785S} and the resonance radii of the bar and spiral arms are unknown in the Milky Way.
Therefore, it is required to trace the kinematics of stars on both sides of the spiral arm at a large range of radius from the bar to the
outer disk, i.e. galactocentric radius range of $0<R_{\rm G}\lesssim10$~kpc.  Also, radial migration is efficient for kinematically colder
stars \citep{2012MNRAS.422.1363S} which can leave imprints in some of the observed chemodynamical relations \citep{Minchev_2014}, and hence
it is crucial to trace stellar kinematics (to be complemented by chemical information from ground based facilities) close to the Galactic
disk midplane.  Gaia has sufficient astrometric accuracy to observe such motions, especially  when traced with bright stars such as Cepheids
\citep{2001A&A...369..339P}. However, in the inner disk and the far side of the spiral arms extinction is a serious problem, and studies at
visible wavelengths, such as with Gaia, are unable to see through the dust to observe any but the intrinsically brightest and/or nearest
stars. A new NIR astrometry mission would allow us to probe this region and determine proper motions and parallaxes of the stars there. This
is the only way that we will be able to get a full 3D picture of the dynamics of this vitally important region of the Galaxy \citep[see][for
a review]{2016arXiv160207702B}.

\subsection{The bulge/bar}\label{sec:nir:bulgebar}
The centre of the Galaxy is dominated by a stellar bar, which has created a peanut-shaped pseudo-bulge \citep{2010ApJ...721L..28N, 2016PASA...33...25Z}. There
are also claims that the bar extends as far as 5kpc from the Galactic centre in the plane of the Galaxy \citep{2015MNRAS.450.4050W}.
Prior to Gaia DR2, dynamical studies of the bar \citep[e.g.][]{2015MNRAS.448..713P} were typically limited to using line-of-sight velocities 
for stars in a relatively small number of fields. 
For a review on the bulge/bar before GaiaDR2 see \cite{doi:10.1146/annurev-astro-081817-051826}.
Progress has been made since Gaia DR2 by using Gaia positions and proper motions to tie the 
relative proper motions found for stars in the VVV Infrared Astrometric Catalogue (VIRAC) as found by \cite{Smith2018} to an absolute rest frame. Using these 
proper motions it has been possible to investigate the kinematics
of the bar \citep{Clarke2019,Sanders2019a}, and even to measure the pattern speed of its rotation \citep{Sanders2019b}. These studies are still
hampered by the relatively low precision of VIRAC proper motions, and the difficulty of relating them to the absolute rest frame.
The pattern speed of the bar is also measured with the combination of Gaia~DR2 and APOGEE by \citet{2019arXiv190511404B}. However, their study is also not free from systematics due to challenges of obtaining accurate photometric distance and small number of Gaia stars in the disk plane.
Indeed, when combining Gaia parallax information with APOGEE spectroscopic information plus other photometric bands, the Bayesian code StarHorse \citep{2018MNRAS.476.2556Q} is able to obtain distances in the bar/bulge area which still show uncertainties of the order of 1~kpc ($\sim$~10\% precision) which is clearly not enough to disentangle the mix of populations in the the bulge area (Queiroz et al. in prep).
A new NIR astrometry mission would directly provide proper motions and parallaxes for stars in this complicated region and allow us to disentangle its dynamics. 
For example, it would allow us to learn about the growth of the bar by observing stars that it recently captured \citep{Aumer2015}
and to what extent a spheroidal component is also present.
%and it would determine once and for all the structure of the inner Milky Way.
Astrometric mapping of the Galactic plane would allow us to answer once and for all critical questions regarding the structure of the inner Milky Way:
is there a large scale, in-plane inner disc or ring \citep{10.1093/mnras/stx2709}; what is the nature of the density and velocity dispersion peak in the central 
degree of the Galaxy \citep{2016A&A...587L...6V}; and how does the apparent multiple component nature of the bulge extend to the innermost regions \citep{doi:10.1146/annurev-astro-081817-051826}?

Cosmological models find that the density profiles of dark matter haloes peak at their centre with $\rho\propto r^{-1}$. However, this is 
in simulations assuming that dark matter is dynamically cold, and without any of the complicated physics associated with the baryonic components 
of galaxies. It has been shown that feedback from star formation can affect the dark matter density profile \citep{2012MNRAS.422.1231G}, and 
the effect of the bar on the dark matter halo is also the subject of debate \citep{2005MNRAS.363..991H}.

The Milky Way presents our best opportunity to study the dark matter content near the centre of an $L^\ast$ galaxy \citep{2005ApJ...627L..89C}, 
which may teach us about the nature of dark matter and/or the processes associated with galaxy formation and evolution on these scales. 
Without an astrometric mission that operates in the NIR, proper motions for these stars will never be determined, and the unique opportunity 
we have to study all-sky and the 6D phase space distribution of stars in the centre of an $L^\ast$ galaxy will be squandered.
However, many of these objects will not have been observed by Gaia so that a new NIR mission will give unique measurements of them with similar accuracy to Gaia. 
The parallaxes, and proper motions at the Galactic centre distances will thus not be very accurate for many new single objects. However, double epoch observations 
for bright stars will give a smaller sample of very accurate parallaxes and proper motions. Most ground based 
IR surveys, for example GRAVITY \citep{2011Msngr.143...16E}, will cover very small patches on the sky compared to the very unique all-sky 
astrometry that a new NIR mission can offer. While GRAVITY will be crucial to investigate better the orbits of stars around the super 
massive black hole at the Galactic centre, a new NIR astrometry mission will constrain the very detailed dynamical and orbital structure of 
many more stars and at greater distances around the black hole (i.e. at larger scales).

\subsection{The Milky Way disc}\label{sec:nir:disc}

Gaia data has uncovered a wealth of unexpected structure and disequilibrium in the disc of the Milky Way \citep{GaiaKatz2018,Antoja2018,2018MNRAS.479L.108K}, which has changed how we view it as a system. It is no longer sufficient to approximate it as an equilibrium structure even in the solar neighbourhood \citep[e.g.][]{Piffl2014}, and we are forced to consider the fact that the local velocity field is rich in structure \citep{GaiaKatz2018}. This includes a `phase-spiral' when looking at the vertical velocity as a function of height above the galactic plane \citep{Antoja2018}, thought to be the consequence of the passage of a dwarf galaxy through the plane of the Milky Way. Disequilibrium associated with the Milky Way's spiral structure affects the velocity field of the disk stars in a way that varies substantially across the Galaxy. The Galactic plane itself shows clear signs of a warp \citep{Poggio2018,Friske2019}. This means that approximations such as thinking of the galaxy as having a simple rotation curve which is a function of radius must be discarded, and we must map the velocity field across the galactic plane. A future astrometric mission will allow us to do this far more accurately than Gaia alone, because of the improved proper motions, and working in the infrared will alleviate the selection effects caused by dust extinction which mean that stars in the midplane of the Galaxy are not seen by Gaia. For example, the phase spiral feature has a scale of 0.1~kpc in distance and a few km~s$^{-1}$ in velocity. To identify the origin of such phase spiral features, it is required to measure this feature at a large radial ($0<R_{\rm G}<\sim20$~kpc) and azimuthal angle ($\theta<\pm60^{\circ}$) ranges with a similar accuracy and number of stars to the currently achieved with Gaia~DR2 in the solar neighbourhood. This will enable to measure the vertical and in-plane oscillation of the Galactic disk in a large radial range, which will be crucial information for Galactoseismology, to examine the influences of the satellite galaxies interactions and the bar formation as well as measuring the stellar disk and dark matter density \citep[e.g.][]{2017Galax...5...44J}.

\subsection{The spiral arms}\label{sec:nir:spiralarms}

It is surprising that very little is known about the spiral structure of the Milky Way \citep[e.g.][]{2017AstRv..13..113V, 2017NewAR..79...49V}. 
The Bar and Spiral Structure Legacy (BeSSeL) survey\footnote{See this \href{http://bessel.vlbi-astrometry.org}{link}.} and the Japanese VLBI 
Exploration of Radio Astrometry (VERA) survey\footnote{See this \href{http://veraserver.mtk.nao.ac.jp}{link}.} have yielded over 100 parallax 
measurements in the spiral arms and the central bar with typical parallax accuracy of about $\pm$20 $\mu$as, and some as good as $\pm$5 $\mu$as. 
These radio measurements are providing good constraints on the fundamental parameters of the Galaxy, including the distance to the Galactic 
centre. Gaia on the other hand is very limited by extinction at visible wavelengths and will not be able to freely probe the Galactic plane \citep{2014ApJ...783..130R}.

The Milky Way is the only galaxy in which we can accurately observe the motion of the stars inside and around the spiral arms. This provides the crucial information for the origin of the spiral arms, which is one of the fundamental questions in the galactic astronomy, but still hotly debated. The classical density wave theory \citep{1964ApJ...140..646L} is recently challenged by transient spiral arm scenario seen in N-body simulations \citep[e.g.][]{2011MNRAS.410.1637S}. Using Gaia~DR1, \citet{2018ApJ...853L..23B} studied the stellar motions in the Perseus arm using 77 Cepheids, and found about $\sim10$~km~s$^{-1}$ level of velocity structures and a tentative sign of the disruption of the Perseus arm. Firmly testing the spiral arm scenario requires to measure the stellar motion inside and around the different spiral arms at different radii with many more tracer samples with a few km~s$^{-1}$ accuracy in velocity and sub kpc accuracy in distance, to resolve the motion at the different sides of the spiral arms.  A new NIR astrometry mission can provide many more samples of stars in the disk plane with enough astrometric accuracy up to about the distance of 8~kpc, and can uncover the stellar motion around the Outer, Perseus, Local, Sagittarius, and Scutum-Centauras arms over large range of Galactocentric radii and azimuthal angles. This will provide an ultimate answer for the origin of the spiral arms.
	
A new NIR astrometry mission will provide a more complete sample of bright stars up to a larger distance, which will enable us to map the
stellar density distribution and identify the over-density of stars for the nearby spiral arms, such as the Perseus, Local and the
Sagittarius arms. \cite{2019arXiv190703763M} identified the stellar over-density of the Local arm, and found a tentative offset from the
high-mass star forming regions identified with VLBI.  Measurements of such offsets for different spiral arms at different radii help to
constrain the spiral arm scenario, and also answer which spiral arms in the Milky Way are the major stellar arms
\citep[e.g.][]{2005ApJ...630L.149B}. In addition, the combination of the accurate density distribution and kinematic measurements for tracer
populations allows us to measure the pattern speed of the spiral arm at different radii by applying the continuity equation locally
\citep{1984ApJ...282L...5T, 2009ApJ...702..277M}.

Spiral arms are the main areas of star formation in the Milky Way, and are responsible for a significant portion of radial migration and disk 
heating. A greater understanding of their structure will help in understanding these processes.
A new NIR astrometry mission would be able to see through dust and would allow us to better map the spiral 
arms and the star forming regions, not just for a few hundred objects as is the case for radio astrometry, but potentially for hundreds 
of millions of objects, revealing much more detailed structure that radio surveys cannot hope to achieve. A new NIR astrometry mission 
could provide model-independent distances and proper motions, avoiding the need to use extinction maps together 
with galaxy modelling but also better constraining extinction maps. Galactic archaeology relies on accurate distances and needs better accuracy towards the spiral arms and the 
galactic centre to give accurate stellar ages and astrophysical parameters in these important regions.

\subsection{Galactic rotation curve and Dark Matter}\label{sec:nir:rotation}
Currently, the gravitational force at different points in the inner disk of the Milky Way is not well known.  This property is key
to learning about the distribution of dark (and luminous) matter near the centre of the Milky Way. A new NIR astrometry mission 
will be the only instrument able to solve this problem since it will unveil the inner dynamics of the disk, in
unprecedented detail, from hundreds of millions of stars, in the bulge, bar, spiral arms, and between the spiral arms. This will
allow us to perform high spatial resolution mapping of the dark matter distribution in these regions. This will be key to
resolving questions regarding the nature of dark matter particles, by showing us whether the Galaxy has a cored or cusped dark
matter halo, whether there are any thin, disc-like components to the dark matter distribution, and whether spiral arms have their
own dark matter component.

In external galaxies, and the outer parts of our own (where we can assume axisymmetry), it is common to approximate 
the gravitational force in the plane using a rotation curve. In the inner Milky Way, this is not an appropriate 
approach because of the non-axisymmetry \citep{2017ApJ...839...61T, 2019MNRAS.482...40K} and because some key tracers (H$_{\rm I}$ or CO regions) provide poor 
spatial coverage \citep{2015A&A...578A..14C}. A more sophisticated approach is required\citep[e.g.][]{2014MNRAS.443.2112H,2015MNRAS.448..713P}
along with more data. VLBI cm measurements of masers \citep{2014ApJ...783..130R} are excellent probes of the rotation curve, 
kinematics and structure in the low latitude regions of the Milky Way. Unfortunately they are too limited in the number 
of targets (at most a few hundreds when their survey is complete), and are only sensitive in star forming regions (thus only 
in the spiral arms or along the dust lanes of the bar), to provide high spatial resolution. VLBI mm measurements with 
ALMA will also play a role in a very near future to do similar science as cm-VLBI, likely with many more detections 
because there are more emission lines at mm- than at cm-wavelengths. Gaia and VLBI will help in the quest to understand 
the inner dynamics of the Galaxy but a NIR capable astrometric mission would allow much greater insight.

The incredible strength of a new NIR astrometry mission with respect to radio interferometry is that it will not be restricted to any particular disk
perturbations or regions and will observe millions of sources. A new mission will be the only way to probe both the low latitude arm
and inter-arm regions, as well as directly inside the bar. It will improve greatly on cm- and mm-VLBI measurements restricted to
star forming regions only, and 21~cm HI or mm CO measurements which do not allow direct measurements of distances and 
proper motions of gas clouds. A new mission is the only way to determine hundreds of millions of velocities at all points in the
Galactic plane and model their dynamics. \citet{2017ApJ...839...61T} and \citet{2019MNRAS.482...40K} demonstrated that the rotation curve
estimates from stellar kinematics are affected by non-axisymmetric structures, such as spiral arms, and it is important to measure the
rotation curve at different azimuthal angles of the Galaxy. A new NIR astrometry mission will provide accurate enough positions and velocities
for a large number of thin disk stars hidden in the dust as far as $\sim8$~kpc and covers the azimuthal angle of $\theta<\pm60^{\circ}$.
This will make it possible to make a comparison with stellar rotation curves of other galaxies, and the cusp-core controversy (among others)
will then be accurately investigated.

\subsection{Baryonic content of the Galaxy}\label{sec:nir:baryonic}

In section \ref{sec:nir:rotation} we described the role that a new NIR astrometry mission can play in the determination of the rotation curve of
the Galaxy and the dark matter distribution. When using dynamics and kinematics to determine the dark matter
distribution it is crucial to have a good knowledge of the distribution of the baryonic content of the Galaxy. The stellar initial
mass function (IMF), the star formation history (SFH), the star formation rate (SFR) and the density laws are fundamental
to determine the baryonic content of the Galaxy. The SFH contains essential information to understand the Galactic
structure and evolution, including key information of its merger history \citep[e.g.][]{Gilmore2001} being a key topic for the
study of the Galaxy. The SFH and the IMF directly influence the chemical composition of the Galaxy, together with the density laws
they constitute a good description of the stellar content of the Galaxy.   

Faint M dwarfs and brown dwarfs have an important contribution in the total stellar mass density of the Milky Way (e.g. \citealt{McKee2015}). A new NIR astrometry mission will be a key instrument to have a full-sky census of faint M dwarfs and brown dwarfs. The photometry, the positions and the parallaxes of a new NIR astrometry mission will be very important to improve the determination of the IMF in the low-mass stars and sub-stellar regime. Complemented by Gaia data we will be able to cover a large mass range, from the fainter and reddest sub-stellar objects up to the more massive, bluer and brighter stars, covering a wide range of the IMF. Furthermore, as a new NIR astrometry mission will survey plenty of old stars in the Galaxy, the photometry and astrometry for the full sky census of M Dwarfs and brown dwarfs will reduce much of the Star Formation Rate (SFR) and Star Formation History (SFH) uncertainty for the oldest epochs (more than 7 Gyrs ago, e.g. \citealt{Mor2019}). Additionally, the positions and parallaxes of a new NIR astrometry mission will allow to determine the density distribution of brown dwarfs and faintest M dwarfs (and pre-main sequence) much beyond the solar neighbourhood. 

A determination of the SFR and the SFH together with the IMF and density distribution will give us the full picture of the distribution of the stellar content in the Milky Way, which is critical for the determination of the dark matter distribution. A model for the star formation efficiency and complementary observations can be used to estimate both the gas needed to form the stars through the history of the Milky Way and the present gas distribution. 
The caveat remains of the impact of radial migration on these estimates. Therefore, once again, it is crucial to trace the whole disk and innermost areas of the Galaxy in order to be able to model and estimate the impact of radial migration onto our estimates for the SFHs and how we interpret the chemo-kinematic-age relations in the Galaxy, and in particular in the disk (see discussion in \cite{2017AN....338..644M} on how age can be used to solve some of the key Galactic archaeology open questions).

\subsection{An age map of the inner Milky Way}\label{sec:nir:ages}
Our understanding of the formation history of the inner regions of the Milky Way is hampered by our lack of knowledge of the large
scale distribution of its relative stellar ages. This is a purely technical limitation. The magnitude of the Main Sequence Turn
Off (MSTO) is the most age-sensitive tracer that allows, with a combination of optical and NIR photometry, to break the
age-metallicity degeneracy in the colour-magnitude diagram to recover the star formation histories of the bulge stellar population
(see for example \cite{Renzini_2018}). Astrometric information is the most critical ingredient for the success of these studies,
as proper-motions with precision < 0.3 mas/yr are required to separate bulge MSTO stars from the foreground disk. For this reason,
bulge stellar-dating studies have been so far restricted to a few discrete fields with deep enough photometry (mostly with HST) to
reach the magnitude of the MSTO. Most of them have been carried out at high Galactic latitudes to minimise the effects of reddening.

Plans for LSST have been suggested by the community to build on the legacy of Gaia and the VVV survey to perform this type of study across
the bulge \citep{2018arXiv181004422G, 2018arXiv181203124B}. These fields are in very complicated areas for LSST due to the combined effects
of crowding and faint magnitudes. Therefore, even in the best case scenario in which some of these vital inner Galactic areas are covered by
LSST, its field-coverage would certainly be constrained to high Galactic latitudes (|b| $>$ 2.5). A new NIR astrometry mission would break all
of these restrictions, providing homogeneous proper-motions, parallaxes, and NIR magnitudes well within the required precision which, when
coupled with the other multi-band surveys, would allow us to estimate accurate ages and metallicities for millions of stars across all the
components of the inner Milky Way. These will constitute extremely valuable targets to be followed up by the Mosaic MOS at ELT and will also
be complementary to the efforts being made in asteroseismology \citep{2017AN....338..644M}, which right now are still concentrated on the
outer disk (e.g. \cite{2017A&A...600A..70A, 2017A&A...597A..30A}).

\subsection{Clusters}\label{sec:nir:clusters}
%
% star clusters and associations - star formation, stellar ages, IMF, low-mass stars
%
Stars are continuously formed in clusters of tens to thousands and evolve together for a shorter ($\sim100$~Myr) or longer time (a few Gyr)
in associations or open clusters, respectively, depending on whether they are gravitationally bound or unbound. Clusters are often located
in the spiral arms of the Milky Way and are composed of young stars that have recently formed in the disk. The stars belonging to a cluster
have roughly the same age and metallicity and can be used to probe the galactic disk structure and formation rate, as well as to study young
star properties and their formation process \cite[see for example:][]{2015ApJ...812..131K} as well as probing radial migration (see
\citealt{2017A&A...597A..30A} and Casamiquela et al.\ 2019 (submitted)).

More than 2000 open clusters are known today, most of them within a distance of 
$\sim2$~kpc from the Sun \citep{2013A&A...558A..53K}. % Kharchenko et al. (2013)
This is roughly 1\% of the total population of over $10^5$ open clusters expected in the Galaxy 
\citep{2016arXiv160700027M}. % Moraux (2016)
The Gaia mission is extending the census of open clusters, most of which are located at high galactic 
latitudes with low interstellar extinction. More than 2000 open clusters were listed as such in the pre-Gaia era
\citep{2002A&A...389..871D, 2013A&A...558A..53K}.  Using Gaia's precise astrometry and photometry \cite{ 2018A&A...618A..93C}
revealed that some of those clusters are only asterisms and confirmed about 1200 clusters. Looking at groups in a five-dimensional
space (positions, proper motions and parallax), \citep{2018A&A...618A..59C, 2019A&A...624A.126C, 2019A&A...627A..35C,
2019arXiv190706872S} have discovered hundreds of clusters undetected until now to limiting magnitudes of $G\sim 17$--$18$. Future
data releases will extend this limit to $\sim5$~kpc, potentially increasing the number of mapped clusters to
$\sim10$\% of the Milky Way total. 
%You can see the history in http://sci.esa.int/gaia/61153-rethinking-everything-we-thought-we-knew-about-star-clusters/

A mission with 14--20 times better proper motion accuracy for common stars and NIR capabilities would cover half of the Galaxy or 
more, including regions towards the Galactic centre. It would thus enable us to probe a much more diverse range of environments 
for cluster formation, in terms of stellar and gas density, and metallicity (all of which increase towards 
central regions). Cluster properties such as stellar density (cluster size and number of members), dynamics, 
and age as a function of location within the Galaxy will provide strong constraints on models of star formation. 
Furthermore, the range of stellar types will be increased to include low-mass stars (subsolar) whose fluxes peak 
in the NIR. This will allow a better characterisation of the IMF at low masses and its 
dependence on the environment, and will again provide clues on the physics of star-forming processes. 
With an increased astrometric accuracy it will be possible to access the more crowded inner regions of each cluster.

The internal dynamics of embedded clusters are needed to study the small scale structure of the molecular clouds
and shed light on the conditions and physics of the cluster at the epoch of star formation. Likewise the bulk dynamics of these clusters
allow one to probe the large scale structure of the molecular clouds and the kinematics tell us about the state of the gas at the epoch of formation.
A NIR option would allow the dusty star forming regions to be globally surveyed for the first time. These important stellar birth places 
are in the regions where extinction intervenes to make visible observations, such as those of Gaia, difficult. 
A new mission will also help to constrain the distances of any gas clouds whose line-of-sight kinematics are similar to nearby stars
while 3-D dust maps of the ISM need good distances.

Open clusters are excellent tracers of the Galactic potential if their tangential motions and accelerations are accurately measured,
particularly as the cluster distances can be very accurately constrained from astrometry and photometry.  Accurate proper motions from combined missions
would allow us to derive cluster membership more clearly. This is already partially possible with the Gaia data alone but a combined mission separated by 
20 years would allow internal accelerations and the initial mass functions to be studied and allow us to probe much larger volumes of the Galaxy. 

ALMA, ALMA-VLBI, and the SKA will help us to understand the way gas clouds evolve (condense, fragment) and form stars with unprecedented 
detail. However, it will not be possible to perform astrometry on a large-scale with these instruments.
A new NIR astrometry mission is an excellent instrument to complement ALMA, as it will enable us to reveal the position-velocity 
phase space of low mass, newly born stars, etc. A new mission will be a unique opportunity to compare the velocities of these new 
stars to the dynamics of the gas, thus fully understanding the star formation processes.

\subsection{Single and multiple stars}\label{sec:nir:stars}

High-precision astrometry in the NIR will have an impact on a wide variety of stellar physics topics. 
%Here, we highlight star clusters and associations, and binary and multiple systems.
With NIR wavelengths we will be able to reach regions in the Galactic plane and near the centre of the Galaxy 
which are particularly interesting for studies of star formation, as they are the common birthplaces 
of stellar clusters and associations.
%
% binary and multiple systems - neutron stars and black holes
%
For binary stars and multiple systems the measurements of astrometric orbits and distances will make it possible to determine accurate masses 
of the components of binary systems, which are crucial to advancement in several areas of stellar physics.
For a planetary system, the mass of the host star must be known in order to determine the masses of the planets. 
Masses of single stars are determined with the help of stellar evolution models, and have uncertainties of up to 30\%, 
leading to 15\% uncertainty in planet masses (see e.g. the exoplanets.org database). This is particularly true for
low-mass stars, which also seem to be the most-frequent planet hosts \citep{2015ApJ...807...45D}.
Masses of binary stars determined from astrometric orbits together with observables to be compared with 
predictions from stellar models provide the necessary constraints for developing the most realistic models. 

More exotic objects, which will benefit from improved mass estimates from binary orbits, are neutron stars 
and stellar-mass black holes. This science case is similar to those envisaged for the SIM astrometry 
mission \citep[see][and references therein]{2008PASP..120...38U}. 
The equation of state for matter at densities beyond those of nuclei is not known. Several proposals exist, 
with different predictions for neutron star masses, and current mass estimates are compatible with all of those.
Likewise, large uncertainties are associated with current mass estimates for Galactic black holes and hamper 
our understanding of their nature. In addition, dynamics of neutron star or black hole binary systems measured from 
accurate proper motions will allow to constrain their formation mechanism.

%The astrometric shifts in a star's position are exceedingly small, and Hipparcos was only able to measure the orbits for a few dozen bright astrometric binaries \citep{2007ApJS..173..137G}. 
Simulations indicate that, with upcoming data releases, Gaia will vastly improve the detection and characterization of exoplanets, brown dwarfs, white dwarfs, neutron stars, and black holes as hidden companions to more luminous, non-degenerate stars (Andrews et al., in prep). These populations will substantially impact a broad range of fields. For instance, the astrometric detection of even a handful of stellar-mass black holes with main sequence or giant companions will resolve outstanding problems in accretion physics and black hole formation physics \citep{2019ApJ...878L...4B}.
A NIR mission would allow exoplanet or compact object detection around cooler, lower mass stars than those observed by Gaia. 
The astrometric signal for a body of mass $M_b$ orbiting a star of mass $M_*$ goes as $M_b (M_b + M_*)^{-1}$, therefore
astrometric detection becomes easier for lower-mass stars at a given distance. In practice this means that the ``distance horizon'' for detection is
increased (resulting in expanded samples) and systems can be detected to smaller masses (e.g., Neptune-like planets).  A similar
limit will hold for long-period binary stars: if the period is 30~years and both have masses 0.5~$M_\odot$, the semi-major axis
will be 12~AU (slightly larger than Saturn's orbit). Since longer-period binaries have larger astrometric signals, these
binaries will be detectable out to larger distances (Andrews et al., in prep).

Proper motions derived from short missions can be affected by systematic errors due to the motions in unresolved binaries and the error depends on the mass and orbital period of the pair. For instance, unresolved double stars can be detected due to the large residuals in the astrometric solution, which is normally based on linear motion for a single star \citep{2018RNAAS...2b..20E}. Additionally, a comparison of the proper motions from each mission with a joint solution from both missions can improve binary population statistics and reveal the acceleration in the orbit. More detailed non-linear modelling for multi-star systems is then needed for the classification of such binaries and this is greatly enhanced by a new mission and the longer time baseline leading to new discoveries.

At the same time, the precise astrometry from Gaia affords the identification of wide binaries, which are detected as resolved stars with consistent proper motions and parallaxes \citep{2017AJ....153..259O,2017AJ....153..257O,2017MNRAS.472..675A}. Since the stars in such binaries were formed from the same pre-stellar material, such binaries can be used to test the chemical consistency of stars from the same population, and can therefore be used to calibrate chemical tagging as a method of uniquely identifying stellar sub-populations in the Milky Way \citep{2018MNRAS.473.5393A,2019MNRAS.482.5302S,2019ApJ...871...42A}. With precise enough astrometry, these binaries can be used to test gravity in the low-acceleration regime \citep{2018MNRAS.480.1778P, 2018MNRAS.480.2660B}. However, the tightest constraints are placed on binaries with orbital separations in excess of 0.1 pc. At these separations astrometric measurements from Gaia are not yet precise enough to discern between various gravity laws (Chanam{\'e} et al., in prep). With a factor of 14--20 improvement in proper motion measurements resulting from the addition of a NIR mission, the combined Gaia and GaiaNIR data set would provide direct, precise measurements of gravity in the low-acceleration regime.

\subsection{Brown dwarfs}\label{sec:nir:browndwarfs}

A strong limitation of Gaia is that it observes in the visible. As such, it is not sensitive to very red objects and sources in high-extinction regions. 
In spite of it, Gaia has been able to identify over 2000 ultracool dwarfs, a diverse ensemble of sources that include very low-mass stars and brown dwarfs in the solar neighbourhood with spectral type later than M7 and that emit the bulk of their radiation in the $JHK$ bands \citep{2017MNRAS.469..401S}. However, of them only a few hundred have L spectral type ($T_{\rm eff} \sim$ 2200--1300\,K), less than ten have T spectral type ($T_{\rm eff} \sim$ 1300--550\,K), and none have Y spectral type ($T_{\rm eff} \lesssim$ 550\,K). Besides, most early-L dwarfs in the solar neighbourhood are main-sequence stars with masses above the hydrogen-burning limit \citep{2008ApJ...689.1295K}. As a result, only a few dozen nearby brown dwarfs have actually been detected by Gaia \citep{2019MNRAS.485.4423S}. 

A comparable number of more distant, younger brown dwarfs, with mid-M spectral type but over-luminous, have also been detected by
Gaia in open clusters and associations of 3--10\,Myr, at 150--400\,pc, and of low extinction (especially $\sigma$~Orionis
and Upper Scorpius). A new NIR astrometry mission would provide us with homogeneous, accurate, parallactic distances of a
very large sample of both ultracool dwarfs in the solar neighbourhood and brown dwarfs in a larger number of young star-forming
regions and juvenile open clusters (e.g. $\rho$~Ophiuchi, Orion Nebula Cluster, Pleiades) down to, perhaps, the deuterium-burning
mass limit \citep{2018Geosc...8..362C}. Such an unbiased sample of very low-mass stars and substellar objects with masses down to
the planetary regime and with very different ages will offer a magnificent panorama of the bottom of the initial mass function,
which is a topic in astronomy that triggers vigorous discussion \citep[see for example][]{2010ARA&A..48..339B}.

\subsection{White dwarfs}\label{sec:nir:whitedwarfs}

White dwarfs (WD) are the final evolutionary stage of intermediate and low mass stars (about 95\% of all stars end as WD
remnants). Their study provides key information about the late stages of the star's life, and also of the structure and evolution
of the Galaxy because they have an imprinted memory of its history \citep{2001ASPC..245..328I, 2005ApJS..156...47L}. The 
LISA mission will study the Milky Way's structure from the gravitational waves produced by close double WDs throughout the whole
Galaxy, and in combination with their motion information obtained from electromagnetic observations the Milky Way's disc and bulge
masses can be derived \citep{2018arXiv181003938K}. Through comparison of the empirical and theoretical Luminosity Functions (LF)
of WDs one can derive the age of the Galaxy and its star formation rate (see \citealt {2016NewAR..72....1G} for a review). The LF
allows the reconstruction of the IMF. 

The Gaia mission has allowed the construction of the largest sample of WDs so far \citep{2019MNRAS.482.4570G}, and the first almost complete volume-limited sample up to 100 pc \citep{2018MNRAS.480.4505J}. The analysis of their position in the Hertzsprung-Rusell diagram has shown a bifurcation not predicted by the models, which cannot be entirely attributed to the existence of a population of helium-rich atmosphere WDs \citep{2018MNRAS.480.4505J}. In addition, the double WD binary population detected by Gaia is of particular interest to the LISA community, although many more double WD binaries are hidden in the dusty Galactic plane making their multi-messenger electromagnetic characterization more difficult.

In this context, a new NIR astrometry mission will be extremely important. As an example, pure-H and pure-He atmospheres can only be distinguished in the NIR regime \citep{2014A&A...565A..11C}. Especially important for the construction of the empirical LF are the oldest members of the WD populations of thin and thick disks and halo, which were formed by high-mass progenitors and evolved very quickly to the WD stage, being very cool and faint in the present day. A NIR facility will allow the characterization of the cool WD population much better than Gaia in the optical, which has limited capabilities \citep{2014A&A...565A..11C} both in terms of detection and parametrization. Furthermore, a new mission will be able to detect and characterise close WD binaries in the dusty regions of the Milky Way's disk and bulge necessary for LISA studies.

\subsection{Astrometric microlensing}\label{sec:nir:microlensing}

\begin{wrapfigure}{r}{0.5\textwidth}
	\centering
	\includegraphics[width=0.5\textwidth]{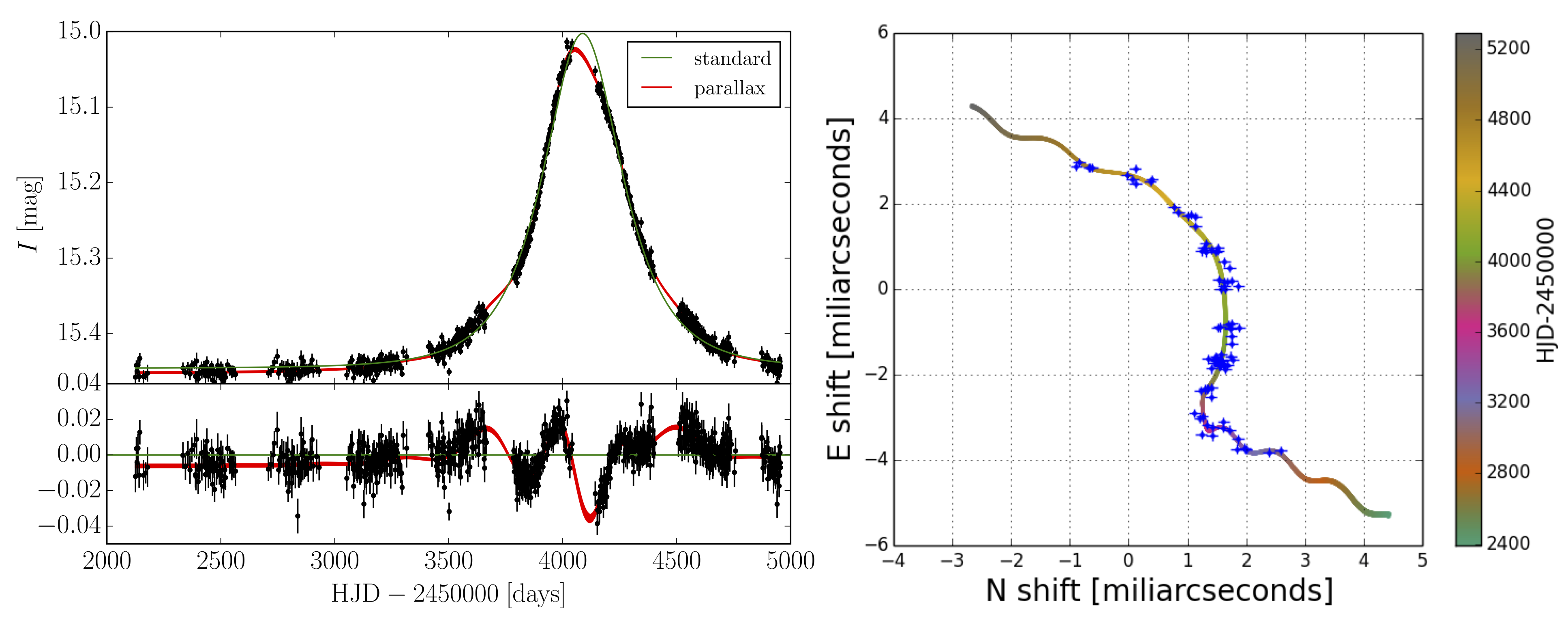}
	\setlength{\belowcaptionskip}{-10pt}
	\caption{\em{Microlensing event, OGLE3-ULENS-PAR-02, a candidate for a $\sim$~10M$_\odot$ single black hole. Left: photometric data from OGLE-III survey from 2001--2008. Right: simulated position of the source (and the centre of light of the two microlensing images) for a similar event, combining source's proper motion, parallax and astrometric microlensing signal. Blue data points are simulated GaiaNIR observations.}}
\end{wrapfigure}

Black holes, neutron stars, brown dwarfs are expected to fill our Galaxy in large numbers, however, such dark objects are very hard to find,
as most of them do not emit any detectable light \citep{2016MNRAS.458.3012W}. About thirty years ago Bohdan Paczy{\'n}ski
\citep{1986ApJ...304....1P} proposed a new method of finding those compact dark objects via photometric gravitational microlensing. This
technique relies on continuous monitoring of millions of stars in order to spot their temporal brightening due to space-time curvature caused
by a presence and motion of a dark massive object.  Microlensing exhibits itself also in astrometry, since the centre of light of both
unresolved images (separated roughly by the 2$\times$ the size of the Einstein Ring, $\sim$~1 mas) changes its position while the relative
brightness of the images changes over the course of the event (e.g. \citealt{2000ApJ...534..213D}).  Astrometric time-series at sub-mas
precision over a range of timescales (depending on the mass and distance of the lens) of several months to years would provide measurement
of the size of the Einstein Ring, which combined with photometric light curve, would directly yield the lens's distance and mass.  So far
(in 2019) there have been thousands of microlensing events discovered in photometry (e.g. OGLE, MOA, Gaia Alerts, future LSST), however, the
mass of the lens has only been measured in a handful of them (e.g. \citealt{2019arXiv190107281W}). 

At the typical brightness of the microlensing events at $I=19$--$20$ mag a new NIR astrometry mission would be capable of providing a good-enough
astrometric signal for the events discovered in photometry by future Bulge or all-sky surveys (including GaiaNIR itself).  The superb
astrometric time series combined with microlensing parallax measurements obtained for events lasting more than $\sim$~4~months, will allow
the mass and distance of the lens to be measured and the nature of the lens to be derived.  A NIR capability would be best for Bulge and
Galactic plane microlensing due to lower extinction and more sources to monitor \citep[see for example][]{2009MNRAS.396.1202K}.

Detection of isolated black holes and a complete census of masses of stellar remnants will for the first time 
allow for a robust verification of theoretical predictions of stellar evolution. Discovery of a large number of stellar-mass black holes in the Milky Way could also help constrain the content of dark matter in form of primordial black holes (e.g., \citealt{ 2016PhRvD..94h3504C}). 
Additionally, it would yield a mass distribution of lensing stars as well as hosts of planets detected via microlensing. 

\subsection{Stellar variability}\label{sec:nir:variability}

%
% stellar standard candles
%
Visible-NIR astrometry will enable extensive local tests of stellar standard candles \citep[e.g.][]{2008PASP..120...38U}. % Unwin et al. (2008), Sect. 8
Accurate distances of Cepheids and RR Lyrae stars throughout the Galactic plane, including the Galactic bulge, will result in
ultimate period-luminosity relations, since the key uncertainties of variable extinction and metallicity are significantly reduced
in the NIR \citep[see Figure 15 in][]{2016A&A...591A...8A}. By extending these measurements to local group galaxies the validity
of these relations can be tested for different chemical environments and galaxy types. 
For Cepheids, a better understanding of the physics of pulsation may improve the applicability 
of the period-luminosity relations for determining accurate galaxy distances. Advancement in pulsation models 
requires accurate masses of Cepheids, which can be obtained from astrometric orbits of binary systems with Cepheid components.
\cite{2011A&A...530A..76W} have shown that the Cepheids are 
very difficult to observe towards the galactic centre (see their Fig. 4) and at low latitudes (see their Fig. 5) and a NIR capable mission would help 
greatly to uncover this difficult region. 
The Cepheids are also crucial for determining the distance scale which is essential for dark energy studies. 
The distance scale will be anchored on Gaia parallaxes soon, but the key uncertainty remains: extinction. NIR photometry will help to 
resolve this problem and significantly contribute to a highly accurate measurement of the Hubble constant, a fundamental quantity for cosmology
\citep{2016ApJ...826...56R}.

Gaia has detected more than 150.000 Long Period Variables (LPVs), mainly Mira and Semi Regular (SR) variable stars \citep{2018A&A...618A..58M}. LPVs are long period pulsating stars in the final stage of their evolution, on the asymptotic giant branch (AGB). Similarly to Cepheids or RR Lyrae, LPV variables show defined period-luminosity relations making them important stellar standard candles \citep{2003ASPC..298..257F}. LPVs also cover a wide range of age, and there is a potential age-period relation \citep{2019MNRAS.483.3022G}. Hence, they can trace the kinematics of different age populations. Moreover, they are much more numerous and brighter than Cepheids, so they could really help in the study of the Galactic structure and extragalactic studies.

The pulsation is accompanied by heavy mass loss which forms a circumstellar envelope of gas and dust which obscures the central
star, absorbing its light and re-emiting it in the IR. The more obscure AGBs (the so-called OH/IR stars), those optically
invisible, are even brighter than Miras and could be even better distance indicators. However, their period-luminosity relation is
barely known due to the dearth of well determined periods \citep{2006A&A...458..533J,2018MNRAS.479.3545B}. Only a NIR mission
would be able to really characterize these stars and to ultimately define their period-luminosity relation. With such a new
mission it would be also possible to study the effect of the different metallicities in the bulge and the disk on the
period-luminosity relation \citep{2015A&A...579A..76J}, so they could be used as a standard candles improving the distance scale.

\subsection{Exoplanetary science}\label{sec:nir:exoplanets}
Around the 2030's space-borne and ground-based exoplanet projects will focus on the characterization of structural and atmospheric
properties of exoplanet systems, particularly those hosting temperate, rocky terrestrial planets amenable to the detection of
biosignature gases indicating the presence of life. However, there will still be uncharted regions of the parameter space
bracketed by mass, orbital separation, and properties of the host stars that can provide critically important constraints on our
understanding of all possible outcomes of planet formation and evolution processes. Precision astrometry is particularly
sensitive to orbiting planets at increasing separation from the host star, and the measurement of orbital motion in the plane of
the sky allows the derivation of the full set of orbital parameters, including the inclination, even for multiple-planet systems (see
\citealt{2001A&A...373L..21S, 2005PASP..117.1021S} and, for a real application, \citealt{2010ApJ...715.1203M}). However, in spite of its
complementarity with other techniques, the applications of precision astrometry to exoplanetary science have thus far been limited
to the characterization of small numbers of known Doppler-detected systems. The required astrometric precision is challenging and
it is only with Gaia that astrometric detection and orbital element fitting for large samples of extrasolar planets are
about to become routine \citep{2018haex.bookE..81S}. Gaia’s large compilation of new, high-accuracy astrometric orbits of
intermediate-separation ($0.5$--$4$ AU) giant planets, unbiased across all spectral types, chemical composition, and age of the
primaries, will allow astrometry to crucially contribute to many aspects of the formation, physical and dynamical evolution of
planetary systems, particularly given its synergy potential with ongoing and planned exoplanet detection and (atmospheric)
characterization programs \citep{2018haex.bookE..81S}. However, even after Gaia’s ultimate exoplanet catalog publication,
key uncharted territories will remain to be explored in the realm of extrasolar planets. 

A new NIR astrometry mission will revolutionize the field of exoplanetary science in two ways. First, it will
establish the presence of planetary-mass companions orbiting classes of stellar and sub-stellar primaries that cannot be
observed with sufficient sensitivity using other techniques and that appear bright in the NIR, enabling a jump of typically a
factor of several in achievable astrometric precision with respect to Gaia. These include two particularly relevant samples:
1) hundreds of ultra-cool early-L through mid-T dwarfs in the Sun's backyard \citep[$d\lesssim 40$ pc, see
e.g.][]{2019ApJS..240...19K}, around which GaiaNIR could complete the census of any existing population of cold giant
planets out to $\sim 4-5$ AU (e.g., \cite{2014MmSAI..85..643S}), with the possibility to access the regime of cold Neptunes and
Super Earths orbiting the nearest, brightest L dwarfs; 2) a statistical sample of maybe thousands of heavily reddened young stars
all the way to the bottom of the main sequence in the nearest star-forming regions ($t\sim3$--$10$~Myr, $d\lesssim
200-300$ pc, see Section \ref{sec:nir:browndwarfs}), around which only GaiaNIR might be sensitive to intermediate-separation
gas giants. These unique contributions of a new NIR astrometry mission would provide a) first-time measurements of the efficiency of planet
formation around brown dwarfs, which is today a matter of hot speculation \citep[e.g.,][]{2012ApJ...761L..20R}, with very little
observational support \citep{2013ApJ...778...38H}, and b) much improved constraints on the true planetary companion mass function
at all separations and at very young ages (when orbital evolution might still be ongoing), in synergy with present and
planned high-contrast imaging programs. All the above results will also greatly benefit from improved parallax estimates for the
host stars, particularly in the low-mass star and ultra-cool dwarf regime (see Sections \ref{sec:nir:browndwarfs} and
\ref{sec:nir:stars}).

Second, systematic GaiaNIR observations of all stellar samples (possibly $\geq10^6$ stars) around which Gaia
(particularly a 10-yr extended mission) is sensitive to planetary companions \citep{2018haex.bookE..81S} would also bring novel
insight on the global architectures of planetary systems. The loss in sensitivity to orbit reconstruction with Gaia beyond
the mission lifetime is well known \citep{2008A&A...482..699C, 2014MNRAS.437..497S, 2014ApJ...797...14P, 2018A&A...614A..30R}. A
new mission launching approximately in, for example, 2040 and using a similar scanning law to Gaia's would allow accessing gas giant companions
with orbital periods in the $\sim30$--$35$~year interval, depending on the new mission's lifetime. Astrometric detection of a planet with a 
non-zero eccentricity can be done with a small fraction of the orbit, so we may be able to determine much longer period planets also.
 Tests show that the presence
of a gap in the data collection does not introduce too large biases in the recovered orbital elements. Aliases have been found to
be easily identifiable, with their presence greatly mitigated when at least one mission is extended. With two short missions
(5+5~year) we expect fractions of false positive and false negative detections under $\sim10$\%; and significantly less for two
long missions (10+10~year). The Gaia+GaiaNIR unbiased census of bright stars screened for gas giants at Saturn-like
distances and the possibility of determining the frequency of multiple systems containing {\em both} giant planets at Jupiter- and
Saturn-like separations would allow, for instance, identification of exact Solar System-like architectures, complementing at
intermediate separations the systematic searches for temperate terrestrial planets around the nearest solar-type stars with
extreme-precision radial velocities \citep[e.g.,][]{2016PASP..128f6001F}, and the characterization of potentially habitable rocky
planets transiting Sun-like stars identified by the PLATO mission \citep{2014ExA....38..249R}.

\section{Improved proper motion science cases}\label{sec:propermotions}

A portion of the rich and violent history of the Milky Way has become clear due to Gaia data. The Milky Way underwent a major merger 
with another galaxy in its early life (around 10 Gyr ago), and the remnant of this merging galaxy makes up the majority of the Milky 
Way's inner stellar halo \citep{Belokurov2018,Helmi2018}. Gaia data has also been used to find the remnants of another significant 
merger in the early life of the Milky Way \citep{Myeong2019}, and to detect and characterise nearby streams of stars, formed when 
satellite galaxies or globular clusters are pulled apart in the potential of the Milky Way \citep{Malhan2018,PriceWhelan2018}. 
However it will not be sufficient to discover and characterise most of the stream-like structures in the halo. Improvements in the accuracy
of proper motions would allow a new mission to resolve tangential motions in streams and local dwarf galaxies, with a potential
accuracy of 2--3$~{\rm km}\,{\rm s}^{-1}$ for specific samples out to $\sim100~{\rm kpc}$. Additionally, improved proper motions will also be 
crucial to help disentangle the mixed populations in the bulge region. 
This is only possible by exploiting the long time baseline allowed by combining Gaia measurements
with those from a future astrometric mission. 

Furthermore, using the Gaia proper motions of tracers in the Milky Way halo, the potential of the Milky Way, and therefore
its mass, was measured out to around $20$~kpc \citep{Posti2019, Watkins2019,Wegg2019}, but this is only thought to be $10$--$20$
percent of its total mass.  The total mass of the Milky Way is an important quantity because it tells us where the
Milky Way fits in a cosmological context. The local group has been called the “Ultimate Deep Field" \citep{Boylan-Kolchin19062016}
because it probes a large co-moving volume at high redshift, and can be observed with great detail. Important questions in
cosmology such as the “Missing satellite" problem and the “Too Big to Fail" problem are related to whether the Milky Way is
typical of galaxies seen in cosmological simulations. To make this comparison in a fair way one needs to know the mass of the
Milky Way.

\subsection{The halo and streams}\label{sec:propermotions:halo}

Streams in the Milky Way halo are formed when satellite galaxies or globular clusters are pulled apart in the tidal potential of the
Milky Way. The stars then drift apart because they are on different orbits and form a (typically thin) band of stars across the sky.
This makes them sensitive probes of the Milky Way’s potential \citep[e.g.][]{Bovy2016,Malhan2019} because the model of their formation is relatively straightforward,
and their narrow distribution across the sky provides tight constraints in some ways (though ones that require more phase-space
information if they are to be useful). This will allow us to determine the dark matter distribution at large radii, including any
flattening of the potential, and the total mass of the Galaxy. A future astrometry mission would provide highly accurate proper
motions and more accurate distances to these stars, which will allow much more precise determination of the potential of the Milky
Way at these large radii.

A further exciting opportunity opened up by improving proper motion measurements for stars in Milky Way streams is that by 
finding ``gaps" in the streams (more easily found if we have proper motions as well as positions for stars), we might see the 
influence of dark matter sub-haloes in the Milky Way's halo \emph{even though they contain no stellar matter}. It is hypothesised 
that such sub-haloes are common in the Milky Way halo, but the only way to detect them is through their gravitational interaction 
with luminous matter. When one of these dark matter haloes has a flyby encounter with a stellar stream, it can create a disturbance 
(or gap) that may be detectable with sufficiently accurate proper motion measurements \citep{2016PhRvL.116l1301B,2012ApJ...748...20C,2016MNRAS.457.3817S}.
The dynamically cold nature of the stellar streams makes them ideal candidates for detecting these flybys.

Another method of determining the mass of the Milky Way out to large radii is by modelling the motion of a tracer population in the
outer halo that can be assumed to be in equilibrium. The major difficulty with this endeavour has historically been that there is a 
degeneracy between the mass enclosed within a given radius and the orbital anisotropy of the tracer population \citep{2006MNRAS.369.1688D}. 
This is only a problem because the only available measured velocity component was the line-of-sight velocity, and for stars in the outer 
halo this component is nearly identical to the Galactocentric radial velocity. The availability of accurate proper motions for these 
stars will break this degeneracy and enable us to learn the mass of the Milky Way from the tracers we see.

Further, it is an open question what fraction of the stellar halo is in substructures and what fraction is in a smooth component, and how much impact the LMC made \citep{2019MNRAS.487.2685E}.  Sometimes 
substructure can be found as an over-density on the sky, but the availability of accurate proper motions will make it easier to find 
substructure in the larger phase-space and even the dark matter wake due to the LMC \citep{2019arXiv190205089G,2019MNRAS.tmpL.102B}.

All of the above questions cannot be answered by Gaia alone due to the low accuracy of the proper motions at great distances ($\sim$~10~kpc).
However, a new mission combined with the older Gaia/Hipparcos/Tycho-2 catalogues would give the much longer baseline needed to get very 
accurate proper motions; remember a factor of 14--20 better in the two proper motion components (see section \ref{sec:sciencecase:measurement}) 
to open up these critical areas to understand the Galaxy in a much grater volume than is currently possible.

\subsection{Hyper-velocity stars}\label{sec:propermotions:hypervel}

\cite{2015ARA&A..53...15B} and references therein, have shown that the origin of hyper-velocity stars (HVSs) is most 
likely due to gravitational interactions with massive black holes due to their extreme velocities. Future and very accurate
proper motion measurements are a key tool to study these objects. Precise proper motion measurements, due to the largely radial 
trajectories of these stars, combined with radial velocities can provide the three dimensional space velocity of these objects. 
Unfortunately, known HVSs are distant and on largely radial trajectories \citep{2015ARA&A..53...15B}. Some HVSs originate in the 
Galactic centre while others have an origin in the disk but an origin in the Magellanic Clouds or beyond is also possible. 
Proper motion accuracies from Gaia (or combined missions) are needed to reconstruct their trajectories and distinguish between 
the different possible origins.

\cite{2005ApJ...634..344G} showed that precise proper motion measurements of HVSs give significant constraints on the
structure (axis ratios and orientation of triaxial models) of the Galactic halo. Triaxiality of dark matter halos is predicted by
cold dark matter models of galaxy formation and may be used to probe the nature of dark matter. Using data from Gaia and a
new NIR mission, combined with distances and radial velocity measurements, would allow for a factor of $14$--$20$ improvement in the
accuracy of the proper motions to advance our understanding of HVSs and the Galactic potential. Adding the NIR
capability would also allow us to probe more deeply into the Galactic centre and potentially detect small populations of HVSs
closer to their ejection location.

\subsection{Co-moving stars}\label{sec:propermotions:comoving}
Simulations \citep{2019arXiv190210719K}  have demonstrated that co-moving stars, even those that are widely separated ($\sim$10~pc), are mostly co-natal. This picture is further borne out by observations \citep{2019MNRAS.482.5302S, 2018MNRAS.473.5393A}, supporting the idea co-moving stars are "clusters of two." Such objects are of high interest in Galactic archaeology because they are significantly more abundant than their open cluster counterparts. The lack of nearby open clusters for detailed spectroscopic follow-up has always been a significant roadblock to calibrating stellar models. The vastly abundant co-moving (and co-natal) stars have enormous potential in changing the landscape of stellar model calibrations (e.g., spectral models, isochrones). The frequency of co-moving stars is a tell-tale sign of Galactic perturbations (e.g., giant molecular clouds and dark matter substructures) and clustered star formation. Most co-moving stars were formed in a clustered star formation environment and were subsequently disrupted by Galactic perturbations \citep{2019arXiv190402159K}. It is the competition of these two effects which sets the number density of co-moving stars. Modelling the statistics of co-moving stars will thus put stringent constraints on these two subtle properties that are otherwise not directly detectable. A new mission will be an important step forward in finding more of these co-moving pairs.

\subsection{Local group}\label{sec:propermotions:localgroup}

Astrometrically resolving internal dynamics of nearby galaxies, such as M31, dwarf spheroidal galaxies, globular clusters, the Large and Small Magellanic Clouds (LMC, SMC), sets requirements on the accuracy. For example, the LMC has a parallax of 20~$\mu$as and an accuracy of about 10\% is needed, which 
is just within the reach of Gaia. Precise mapping of dark matter (sub-)structure in the local group and beyond is possible with accurate proper motions.
Gaia can only just directly measure internal motions of nearby galaxies. Combining proper motions from two Gaia-like missions
opens up the tantalising possibility of accurately measuring their internal motions and thus resolving the dynamics within the Local Group.
A number of science cases are within reach:
\begin{itemize}
	\item Using photometric distances to the LMC allows us to learn about internal dynamics and structure. 
% More an improved parallax effect	\item Mapping the ISM in the LMC would also give new insight.
	\item Using proper motions to determine rotational parallaxes to M31, M33 and other local galaxies.
	\item Dynamical measurements of the mass distribution of the Milky Way and M31.
	\item Probe the internal kinematics classical dwarf spheroidal galaxies. 
	\item Mapping the dark matter sub-structure throughout the local group out to M31.
	\item Resolve the core/cusp debate\footnote{Discrepancy between observed dark matter density of low-mass galaxies and density profiles predicted by cosmological N-body simulations.} with 6D phase space information.
	\item Ultra-faint satellite galaxies would be accessible and their orbits could be better determined than by Gaia alone \citep{Simon_2018}.
\end{itemize}

Dwarf spheroidal galaxies (dSph) are fascinating systems and likely fossils of the re-ionisation epoch\footnote{This section is extracted from an unpublished note from Rodrigo Ibata.}
In order to study their internal proper motions there is a need to disentangle the Milky Way environment using very accurate 
proper motion measurements.  The dSphs are small, almost spherical, agglomerates of stars that orbit more massive hosts. 
In the local group these dSph galaxies are the most numerous class of galaxies, with 15 discovered around the Milky Way, 25 
around the Andromeda galaxy, and possibly one orbiting M33. Several other nearby galaxies have also been found to harbour 
these satellites. It is believed that hundreds of dSphs and globular clusters were accreted during the formation 
of the Galaxy, and their remnants should be visible as coherent phase-space structures. The challenge 
is of course to identify the stars with common dynamical properties.

An interesting aspect of these galaxies is their identification as possible cosmological “building blocks” of the type predicted 
by cold dark matter theory. A generic feature of a Universe in which the dominant form of matter is in the form of 
cold dark matter is that galaxies should be surrounded by tens of thousands of dark matter satellites 
with masses comparable, or larger, than those of dwarf galaxies. While the number of observed satellites is very much lower than 
this, it has often been suggested (or assumed) that baryons collapsed to form substantial stellar populations in only the most 
massive of the dark satellites. Other dark satellites may have lost their baryons, or had star-formation suppressed for a variety 
of reasons. Thus the dwarf satellite galaxies may represent the surviving remnants of the original population of primordial galaxies 
that coalesced and merged to form our Milky Way.

To confirm this picture we need to understand the internal dynamics of dwarf galaxies, and to date, the necessary observations 
to test these ideas have remained beyond our reach. The issue is that there is a trade-off between dark matter content and tangential 
anisotropy and/or tidal heating. A system with stars on preferentially tangential orbits to the Galaxy and that has also been heated by 
repeated tidal encounters with the Milky Way can masquerade as a highly dark matter dominated galaxy. The only way to be sure of 
the true dark matter content is to measure the internal proper motions of the stars in the dwarf galaxy, not only the bulk proper 
motion of the system (feasible with Gaia). Only the combination of Gaia and GaiaNIR can hope to achieve this challenging goal.

\subsection{Prospects for exoplanets and long period binaries}\label{sec:propermotions:exoplanets}

Proper motions from short missions are affected by systematic errors due to the motions in unresolved binaries and the
error depends on the mass and orbital period of the pair. Such systems are poorly surveyed at present. The unresolved double stars
can be detected due to the large residuals in the astrometric solution, which is normally based on linear motion for a single
star. Additionally, a comparison of the proper motions from each mission with a joint solution from both can improve
binary population statistics and reveal the acceleration in the orbit. More detailed non-linear modelling for multi-star systems
is then needed for the classification of such binaries which is greatly enhanced a new astrometry mission and the longer time baseline,
leading to new discoveries.
 
\cite{2019A&A...623A..72K, 2018ApJS..239...31B} 
have shown that Hipparcos and Gaia astrometry can be used to measure changes in proper motion, i.e., accelerations 
in the plane of the sky, which opens the possibility to identify long period orbital companions otherwise inaccessible 
while also constraining their orbits and masses. 
Synergies between the two Gaia-like missions and the PLATO mission \citep{2014ExA....38..249R} would allow a comprehensive comparison of the short 
period planets detected by PLATO and the long period planets from two Gaia-like missions giving quantitative insight into possible Solar System analogues.

\subsection[]{The Solar System}\label{sec:propermotions:solarsystem}

A strong impact of Gaia and of the proposed follow-up mission GaiaNIR is on the study of the population of small Solar System bodies, asteroids, comets, and planetary satellites. Precise fundamental ephemerides of the planets in the Solar System rely mostly on radar and spacecraft data for the inner Solar System, spacecraft data for Jupiter and Saturn, and lunar laser ranging for the Moon. The link to the ICRF is provided by VLBI observations of the spacecraft \citep{2014arXiv1408.3302H, 2015HiA....16..219F}. Radar techniques have also been exploited on small bodies, such as Earth-crossing asteroids (NEOs) but remote optical observations are still required, as only they are the capable of covering the largest sample of object sizes and distances. 

Infrequent spacecraft flybys of asteroids are very useful for determining masses and other physical properties, but do not always
constrain the orbital parameters very well. So far only 20 small bodies have been visited, while a NIR survey will reach hundreds 
of thousands. For the distant outer planets, the modern astrometric technique is to observe the
positions of the natural satellites and infer the position of the centre-of-mass of the planet-satellite system. Better orbits for
the satellites are also of interest for modelling the dynamical history of the satellite systems. In the inner Solar System, there
is much current interest in near-Earth asteroids (NEOs) that pose some risk of colliding with Earth in the future. These are
generally small, faint objects that do not have long-term (or any) observing histories, as their orbits are closer to
the Sun than the Earth’s orbit making their detection particularly challenging. Estimating the future paths of these objects and
their risk to Earth therefore requires accurate absolute astrometry. Also, to trace back their origin to the Main Belt, the
history of collisional families must be studied. Their spread in the orbital element space can be used as an astronomical clock for
their evolution \citep{2006Icar..182..118V, 2015Icar..257..275S}, and is driven by the Yarkovsky acceleration \citep[recoil force
due to the emission of thermal photons][]{Delbo20081823}. Measuring the Yarkovsky effect, which causes a secular drift in the orbital 
semi-major axis, is one of the main challenges of asteroid science.

Today, Gaia is demonstrating the potential of high accuracy astrometry and photometry in the dynamical and physical characterisation of the population of small Solar System bodies. Among modern challenges that require accurate astrometric measurements and precise orbits are, determining the masses of asteroids, predicting stellar occultations by asteroids \citep{2007A&A...474.1015T}, monitoring the impact risk of potentially hazardous near-Earth objects, which may exhibit very fast apparent motion, and, conversely, the determination of the orbits of very slow and faint bodies, the so-called Trans-Neptunian Objects (TNOs).

Asteroid masses can be determined by very precise astrometry before and after mutual encounters \citep{1996AJ....112.2319H}. However, asteroid-asteroid encounters that are useful in determining asteroid masses are infrequent events for ground-based observers, due to the limitations in the astrometric accuracy. Unknown masses of asteroids contribute “noise” in the ephemeris of the planets in the inner Solar System \citep{2002A&A...384..322S}, although the last decade’s worth of ranging data from Mars spacecraft have allowed for improvements in the determination of the masses of the asteroids that significantly perturb that planet’s orbit \citep{2012IAUJD...7E..38K}. However, Gaia astrometry is accurate enough to directly provide masses of the largest 100 perturbers \citep{2007A&A...472.1017M}.
Gaia astrometry of stars is also being effectively used to improve stellar occultation predictions. For instance, the use of Gaia astrometry in the case of the occultation targets by the TNO 2014 MU69 has permitted positive observations and provided strict constraints on its orbit. This success has been fundamental for planning of the flyby by the New Horizons (NASA) space probe \citep{2018DPS....5050906B}. 

While successful occultations and mass determination are based on the instantaneous knowledge of very accurate orbits, taking into account subtle long-period, or secular effects, requires observations over a long time span. For instance, measuring the Yarkovsky acceleration directly in the Main Belt would give a major contribution to our understanding of asteroid families and to the mechanisms that lead to the injection of asteroids on Earth-crossing trajectories. The best current Yarkovsky determinations are based on the joint use of both Gaia and previous ground-based astrometry spanning several years (or decades). They can be successful only when the orbit uncertainty (on the semi-major axis) reaches a few $\sim$~100~m or less. The low accuracy of pre-Gaia asteroid astrometry is a clear limitation, as the average RMS of the residuals for CCD astrometry in the Minor Planet Centre data base is $\sim$~400~mas \citep{2013A&A...554A..32D}. Careful data selection, and zonal de-biasing strategies for old positions derived relatively to stars from pre-Gaia catalogues, are required \citep{2003Icar..166..248C, 2010Icar..210..158C,  2015Icar..245...94F}.
With this approach, Gaia is expanding the limit of detection for Yarkovsky from the NEO region to the inner Main Belt.

The possibility of a direct, accurate measurement of Yarkovsky deep into the Main Belt on collisional family members, would be a major breakthrough brought about by an astrometric follow-up mission.  The expected improvement factor in the orbit accuracy obtained by combining the data of two astrometric missions on a 20 year baseline, is similar to the improvement estimated for the stellar proper motion, in the case of NEOs, Main Belt asteroids or natural planetary satellites. For them, observations will be distributed all along their orbits. On the other hand, for small bodies whose orbital period is substantially longer than the Gaia mission alone the gain in accuracy is potentially much higher, as in this case it depends strongly on the fraction of the orbital period that is covered. This is the case for objects exterior to the obit of Jupiter such as the Centaurs (periods between 20 and 150 years) or the TNOs. This population also harbours the highest priority target for stellar occultations, the only technique capable of providing accurate size measurements and vertical density profiles of their atmospheres (such as for Pluto) at those distances. 

The second major contribution of a NIR mission would be its complementarity with respect to the spectral coverage of Gaia. In fact, Gaia is building a new taxonomy based on the visible, but asteroid classifications including the NIR breaks the degeneracy of certain spectral types and bring additional information on the composition \citep{2009Icar..202..160D}. For wavelengths exceeding 1 micron, slope changes can occur and wide absorption bands, diagnostics of the mineralogy, can affect the slope at the extreme of the GaiaNIR spectral coverage (1.8-2 microns). 
%As the spectral features are very wide, spectral resolution is not an important issue. On the other hand, the scientific reward is enormous, as our current knowledge of NIR asteroid spectra comes from the order of 1000 objects only. 
Being able to retrieve mineralogy information for 100,000's of objects would be a major milestone with a direct impact on the models describing the formation and the evolution of the Solar System \citep{2014Natur.505..629D}. GaiaNIR would then be the largest NIR-inclusive survey of the asteroid population, providing a new portrait of the asteroid belt composition. 
This constitutes a clear advantage, as physical properties such as surface composition, or rock porosity, directly affect the Yarkovsky effect and are related to the properties of the family parent bodies. Extension in time and wavelength coverage are thus complementary aspects that give a clear strength to GaiaNIR in Solar System science.

\subsection[]{Astrometric radial velocities}\label{sec:propermotions:astrometricrv}	
High-accuracy astrometry permits the determination of not only stellar tangential motion, but also the component along the line-of-sight. Such non-spectroscopic (i.e. astrometric) radial velocities are independent of stellar atmospheric dynamics, spectral complexity and variability, as well as of gravitational redshift. For a detailed discussion see \citep{1999A&A...348.1040D}. There are three methods for measuring astrometric stellar radial velocity a) changing annual parallax, b) changing proper motion (perspective acceleration) and c) changing angular extent of a moving group of stars. All three have significant potential for a combination of Gaia and a new astrometric mission. The baseline design of GaiaNIR does not envisage any spectrograph on-board, but perspective acceleration will allow to measure velocity vectors up to a few ten's of parsecs. This may be relevant for brown dwarfs, for nearby exoplanets ejected from their host star systems and for objects at the outer edges of the Solar system.

\section{Maintenance of the celestial reference frame}\label{sec:refframe}

Celestial Reference Frames (CRFs) represent catalogues of sources with
accurately known positions (possibly as functions of time). CRFs are important for a number of reasons:

\begin{itemize}
	\item To refer positions of fixed or moving sources
	\item To detect tiny motions (e.g. exoplanet detection by relative astrometry)
	\item To quantify the motion of sources without any kind of bias:
	\begin{itemize}
		\item Motion of stars in the Galaxy
		\item Differential rotation of the Galaxy
		\item Dynamics of star clusters
		\item Rotational and translational motion of external galaxies
		\item Dynamics of the Solar System
	\end{itemize}
	\item Source cross-identification in $\gamma$, X-ray, visible, IR and radio wavelengths
	\item To monitor the rotation of the Earth and study tectonic plate motions using ground-based VLBI
	\item Angular positions (and distances) of quasars, galaxies, stars, planets, spacecraft navigation, GNSS maintenance, etc.
\end{itemize}

\begin{wrapfigure}{r}{0.69\textwidth}
	\centering
	\includegraphics[width=0.31\textwidth]{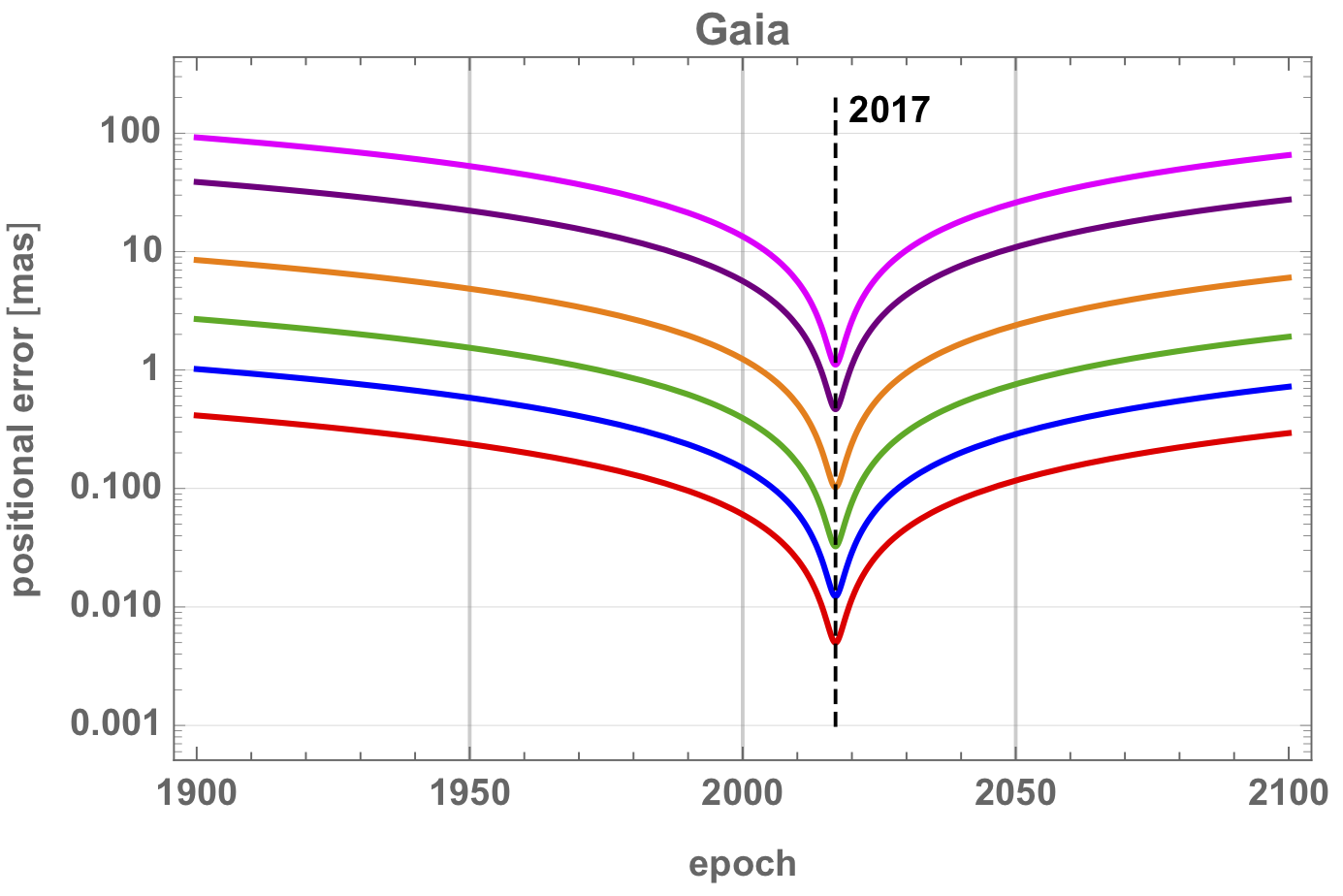}
	\includegraphics[width=0.374\textwidth]{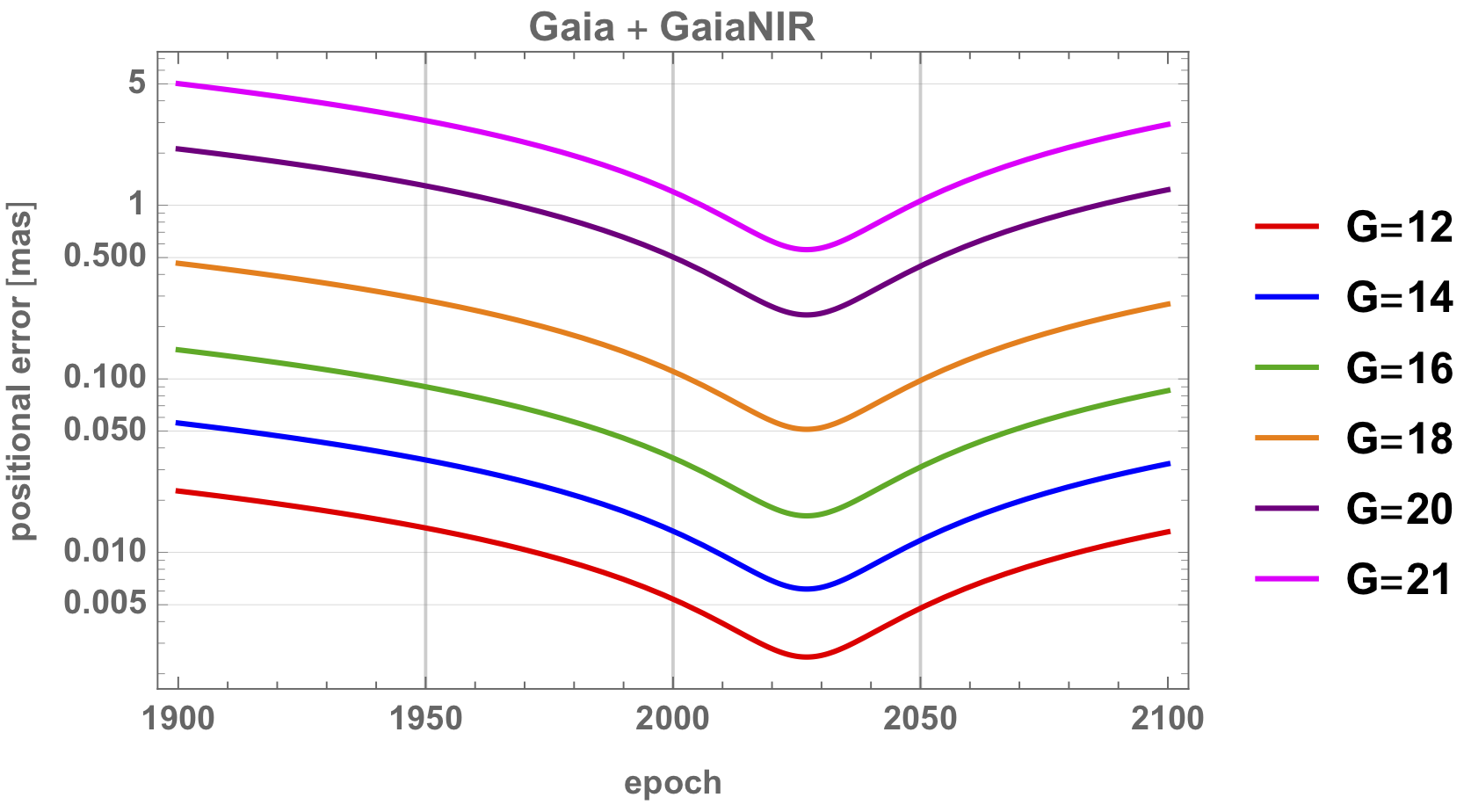}
	\setlength{\belowcaptionskip}{-10pt}
	\caption{\em{Degradation of the astrometric accuracy of the individual sources in the Gaia catalogue (left pane) and of the common
  solution using 10 years of Gaia and 10 years of GaiaNIR data (right pane) depending on the G magnitude and time from the reference epoch
  (J2017 for the Gaia catalogue and a mean epoch of Gaia and GaiaNIR taken here as J2027). Note the very different vertical scales.}} \label{fig:degradation}
\end{wrapfigure}

Gaia is expected to provide the first highly-accurate Celestial
Reference Frame (Gaia-CRF) in the optical based on extragalactic
sources, thus being less sensitive to possible proper motion errors
of the sources.  The uncertainties of typical individual positions of
Gaia-CRF based on 5 years of data is 0.1 mas. This accuracy will
degrade over time because of errors in the spin of the Gaia-CRF and in
the proper motions due to the acceleration of the solar system
barycentre relative to the rest frame of the quasars.  These
uncertainties can be expected at the level of about 0.003 mas/yr, so
that the uncertainty will double within 50 years after the reference
epoch of Gaia.

However, one million quasars expected in the Gaia-CRF represent a tiny
fraction of the Gaia sources. With the typical angular distance
between the quasars of about 6 arcminutes the Gaia-CRF is not dense
enough to provide a suitable reference grid needed for forthcoming
Extreme, Giant and Overwhelming telescopes but also for smaller
instruments currently operating or being planned. Moreover, most of
the quasars of Gaia-CRF are rather faint (between magnitudes $G=19$
and 21) so that the accessibility of the reference frame is given only
in the optical and almost exclusively in that interval of magnitudes.

For this reason the whole Gaia catalogue with its accurate proper
motions of stars is de facto used as a standard CRF. Because of the errors 
in proper motions the accuracy of this
frame degrades with time. The accuracy degradation is almost linear in
time with the slope equal to the error in the proper motions.  This is
depicted on the left pane of Fig. \ref{fig:degradation}. With proper
motions of up to 20 times better than for Gaia, the degradation of the
common reference frame becomes much slower (right pane of
Fig. \ref{fig:degradation}). Due to the observational principles of
Gaia there are six degrees of freedom the orientation $\vec{\epsilon}
= (\epsilon_x,\ \epsilon_y,\ \epsilon_z)$ and the spin $\vec{\omega} =
(\omega_x,\ \omega_y,\ \omega_z)$ of the resulting reference
frame. The orientation is determined using a set of a few thousand
radio ICRF3/VLBI sources with known positions and proper motions while
the spin is determined using a set consisting of hundreds of thousands
of quasar-like objects identified in the Gaia catalogue using external
photometric surveys of quasars.

The need for a global survey mission like Gaia to maintain the
realisation of the CRF is a science objective in
itself as it lies at the heart of fundamental astrometry and provides
a reference grid for much of modern observational astronomy.  An
important aspect of reference frames is to link them, cross-matching
with absolute coordinates, to reference frames at other wavelengths to
produce reference grids for various surveys. This requires the
maintenance of the accuracy of the Gaia visual reference frame at an
appropriate density that is useful for new surveys. A key improvement
is to expand the Gaia visual reference to the NIR, increase its
density in obscured regions and to then link this new reference frame
to the ICRF. Maintaining the visible-NIR reference frame is a vital
service to the community and is not feasible from the ground.

Experience in comparing astrometry at difference wavelengths has
been gained from cross matching the optical Gaia-CRF and the radio
ICRF2/ICRF3. The so-called radio-optical offsets have been detected and
are being extensively discussed in the literature \citep{2016A&A...589A..71B,2019ApJ...873..132M}.
Although these offsets are very important for the discussion of
individual objects, they are significant for a small fraction
of sources and also uncorrelated from one source to another. This is
therefore only an additional source of noise. One can also expect
that quasars have astrometric variability at the 10~$\mu$as level due to
internal structure and activity. 

%------------------------------------------------------------------------

Eventually, the quasars will be found from the observational data of
Gaia itself using quasar classification and this will also hold for
future missions. \cite{2008AIPC.1082....3B}
developed a method using Gaia photometry by low-dispersion spectra and
showed with simulated data, that it is possible to achieve a pure
sample of quasars (upper limit on contamination of 1 in 40000) with a
completeness of 65\% at G = 18.5, and 50\% at G = 20.0,
even when quasars have a frequency of only 1 in every 2000 objects.

%The efficiency of photometric detection in the NIR can be expected to improve.

%------------------------------------------------------------------------

With a different approach \cite{2015A&A...578A..91H} aimed to discover
quasars using only their lack of proper
motions.  The technique has been further validated after the release
of the Gaia DR2 data \citep{2018A&A...615L...8H}.  Astrometric
selection of extragalactic point sources also has the potential to
select other extragalactic point sources, e.g. potentially new classes
of objects.

It is well known that some quasars may have non-zero apparent proper
motions and parallaxes (the latter because of time-dependent proper
motions projected on the 1-year parallax pattern) due to the quasars
internal structure \citep{2011A&A...526A..25T}, so that only a part of
quasars can be found using purely astrometric approach. Nevertheless,
purely astrometric selection is attractive since it is unbiased by
assumptions on spectra and photometry.  The incompleteness of quasar
samples based on selection by visible photometry has been studied
intensively for many years and it is now well established that such
samples miss a substantial number of, in particular dust-reddened,
quasars (see, e.g., \cite{2013ApJS..204....6F,2016ApJ...832...49K} for
recent studies).  This is a serious problem for a number of important
issues in astrophysics starting with the study of the quasar phenomena
itself.  For example, Broad Absorption Line quasars are systematically
underrepresented in quasar samples.

The results of \citet{2015A&A...578A..91H} show that with the expected
accuracy of Gaia astrometric selection gives 20\%-50\% of stellar
contaminations depending on the galactic latitude. In the set of
sources observed by both Gaia and GaiaNIR, due to more accurate proper
motions one should expect a truly dramatic improvement: the percentage
of stellar contaminants will be reduced by a factor of up to 400, thus
leading to an extremely clean quasar sample everywhere on the sky.

%-----------------------------------------------------------------------------

In conclusion, a new NIR astrometry mission will be necessary in the coming
decades in order to maintain the visible realization of the CRF. This 
on its own is a strong science case due to the
need to maintain dense and accurate reference grids for observational
astronomy in the visible. However, the addition of NIR astrometry will
increase the density of this grid in obscured regions and provide a
link to the ICRF in a new wavelength region and drastically improve the quality of
the Celestial Reference Frame.

\subsection{Proper motion patterns, real time cosmology and fundamental physics}\label{sec:refframe:patterns}

A number of specific proper motion pattern of quasars are expected and
represent a scientifically exiting topic by themselves. First sort of
systematic pattern in proper motions of quasars comes from the
acceleration of the Solar system relative to the rest frame of quasars
causing a drift of secular aberration
\citep{Fanselow1983,1995ESASP.379...99B,1997ApJ...485...87G,1998RvMP...70.1393S, 2003A&A...404..743K, 2003AJ....125.1580K,2006AJ....131.1471K, 2014MNRAS.445..845M, 2016A&A...589A..71B}.
This acceleration is dominated by the acceleration from the motion of
the Solar system in the Galaxy. The amplitude of the corresponding
proper motions is $\sim$4.3~$\mu$as~yr$^{-1}$. Already with Gaia data,
this acceleration can be measured with an accuracy of
$<0.5~\mu$as~yr$^{-1}$ meaning a 10-sigma detection of the effect.
From a combined optical-NIR solution the effect will be measured with
2 or 3 digits giving both much better stability of the resulting
reference frame and a stringent constraint for the modeling
of the galactic dynamics.

%Fanselow1983
%Fanselow, J. L. 1983, Observation Model and parameter partial for the JPL VLBI
%parameter Estimation Software MASTERFIT-V1.0, Tech. Rep.

Another pattern in the proper motions of quasars comes from
the instantaneous velocity of the Solar System with respect to the
cosmological microwave background (CMB).  This will cause extragalactic sources
to undergo an apparent systematic proper motion. The effect is referred to as cosmological or 
parallactic proper motion \citep{1986AZh....63..845K}. 
This effect depends on the redshift, $z$, and fundamental 
cosmological parameters can in principle be determined from the motion. The velocity of the Solar System with respect to 
the rest frame of the observed Universe can be measured from a dipole pattern in the CMB temperature.
Observations of the CMB indicate  that $v = 369 \pm 0.9$ km~s$^{-1}$ in the apex direction with Galactic longitude $l = 263.99^{\circ}$ and latitude $b = 48.26^{\circ}$ 
\citep{2009ApJS..180..225H,2014A&A...571A..27P}. This motion should produce a parallactic shift of all extragalactic objects towards the antapex.

Both the acceleration of the Solar System and the cosmological proper
motion give rise to dipole patterns in the proper motions of distant
objects.  However, the former does not depend on the redshift, while
the latter does, which makes it possible to separate the effects.
\cite{2016A&A...589A..71B} showed that the cosmological proper motion
effect is only just within the theoretical limits of the Gaia mission
but the centroiding of extended Galaxies may result in the effect
being too difficult to measure. However, improving the proper motions
by a factor of 14--20 with a new astrometry mission 20 years later would bring
such measurements safely within reach. Thus, at
least for quasars with $z<1$ the Hubble law can be verified by the
model-independent trigonometric parallaxes.

There are additional, rather speculative scenarios leading to
additional pattern of proper motions in certain
cosmologies. \cite{1966ApJ...143..379K} pointed out that an
anisotropic Hubble expansion of the Universe would result in a certain
quadrupole pattern of proper motions of quasars that could be a
function of the redshifts and can be measured in
principle. \cite{2009PhRvD..80f3527Q,2012PhR...521...95Q} used the
term cosmic parallax for the varying angular separation between any
two distant sources, again caused by the anisotropic expansion of the
Universe. Some model predict that anisotropy of the expansion cannot
be constrained by the CMB measurements. The predicted systematic
proper motions are at the level of 0.02--0.2~$\mu$as~yr$^{-1}$ are are
within the reach of the common Gaia--GaiaNIR astrometry that will be
used to confirm or constrain these scenarios.

Finally, one more source of systematic proper motions is gravitational
waves. It is well known that a gravitational wave passing through an
astrometric observer causes shifts of observed position of all sources
of the sky. Those shifts depend on the mutual orientation of the
direction towards the source and the propagation direction of the
gravitational wave as well as on time. This effect allows for an
astrometric detection or constraints of gravitational waves
\citep{1996ApJ...465..566P,1997ApJ...485...87G,2011APS..APRG12004B,2018CQGra..35d5005K}.
%Because of the nature of astrometric solution that linear proper motions 
Two regimes for astrometric detection of gravitational waves
should be distinguished \citep{2018CQGra..35d5005K}. The gravitational
waves with periods longer than the time span covered by astrometric
data (i.e., many years; these are primordial gravitational
waves of cosmological nature) are absorbed by proper motions. One can
show that the corresponding systematic proper motions of quasars are
dominated by the quadrupole. Since the effect is indistinguishable of
residual systematic errors in the solution, this allows only to
give an upper estimate of the energy flux of primordial gravitational
waves with all frequencies below a certain limit
\citep{1997ApJ...485...87G,2012A&A...547A..59M} The current estimate from VLBI astrometry
based on the data of \citet{2011A&A...529A..91T} is
$\Omega<0.009\times H/100 {\rm km/s/Mpc}$ for frequencies below 1.5
nHz \citep{2012A&A...547A..59M}. Gaia is expected to improve this
limit by a factor of 80. The combined Gaia--GaiaNIR solution can
improve Gaia results by a factor of up to 400.

Another regime is the detection of higher-frequency gravitational
waves from binary supermassive black holes. Recently, an algorithm for
the detection was suggested \citep{2018CQGra..35d5005K} and is now
being implemented for Gaia data. A new mission will give another improved
opportunity to detect the signature of gravitational waves
from astrometry. Compared to Gaia, the detection limit of GaiaNIR is
improved both because of larger number of observed sources and the
possibility to use drastically more accurate proper motions for the
sources common to both missions.

Some additional tests of fundamental physics can be performed with GaiaNIR data
which both provide longer observational time span (combined with Gaia) and
allow for accurate astrometry in optically obscured regions. A few examples here are:
\begin{itemize}[itemsep=0.1pt,parsep=0pt]
	\item More accurate tests using the motion of asteroids in the weak field limit of the Solar System (PPN-$\gamma$ \& PPN-$\beta$, time dependence of the gravitational constant). %, alternative theories of gravity, possible quantum effects).
	\item Strong field gravity tests at Galactic centre in the NIR (this includes both deflection of light in the gravitational field of the central Black Hole and the dynamics of the stars near the Galactic centre).
	%\item Probe the strong gravity with binary systems including neutron star or black hole binaries and learn the mass, equation of state, accretion discs in the strong field limit.
	%\item Weighing neutron stars independently.
\end{itemize}

\begin{wrapfigure}{r}{0.35\textwidth}
	\centering
	\includegraphics[scale=0.35,trim={0cm 2.6cm 28cm 0.4cm},clip]{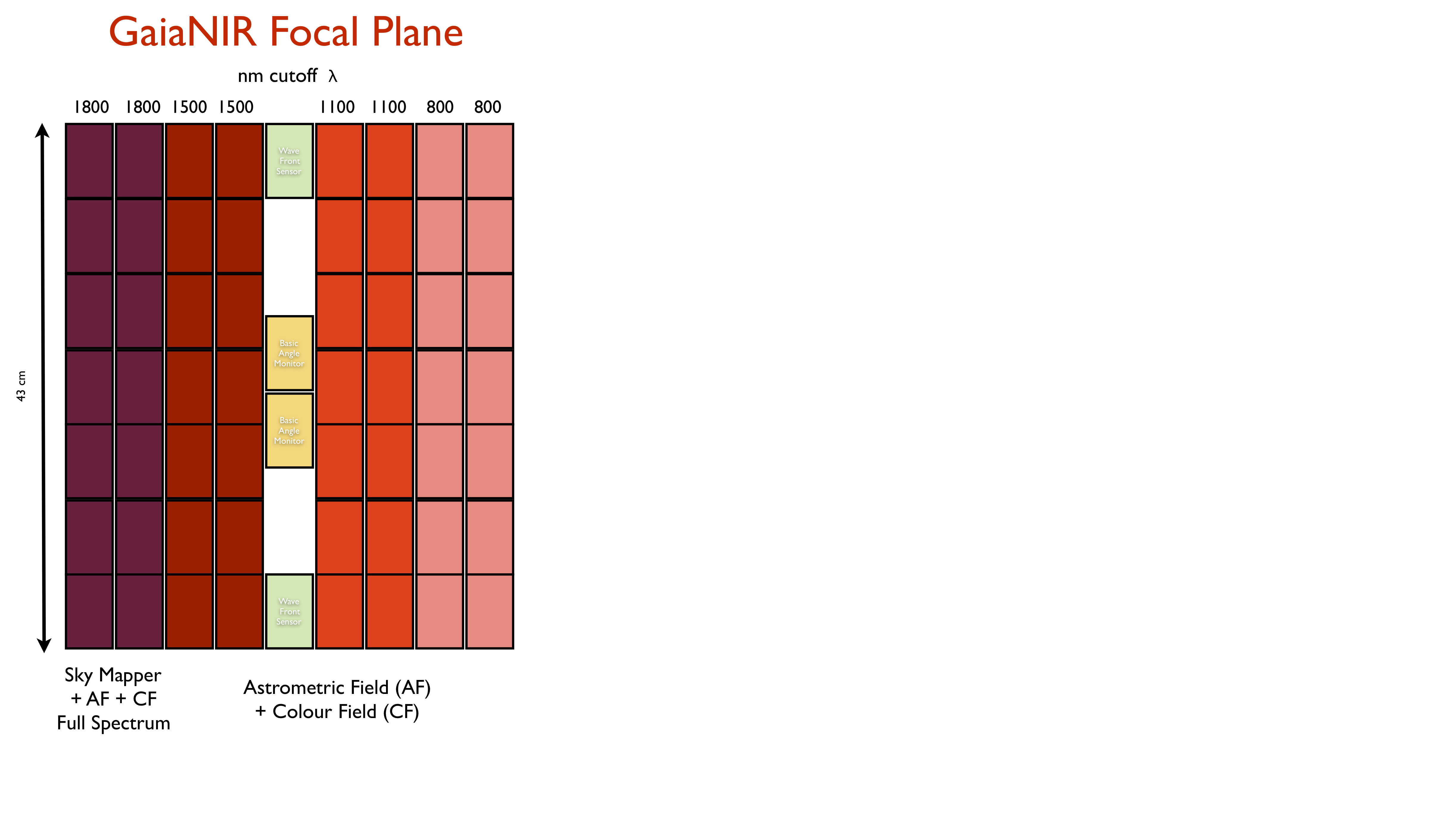}
	\setlength{\belowcaptionskip}{-20pt}
	\caption[]{\em{Proposed focal plane array and filter bands used in the GaiaNIR CDF study (\href{http://sci.esa.int/future-missions-department/60028-cdf-study-report-gaianir/}{see page 203}). The array consists of 60 NIR detectors, arranged in 7 across-scan rows and 9 along-scan strips (out of which 8 are for the astrometric/photometric field, divided into 4 photometric fields (i.e. 4 different cut-off wavelengths each starting from $\sim$~400~nm). The new array is less than half the size of Gaia's.}}\label{fig:bands}
\end{wrapfigure}
\section{Photometry}\label{sec:photometry}

For a new NIR astrometry mission it is proposed that the photometry be implemented as filters deposited directly on the detector material forming 
4 broad bands. This will provide photometry of all sources, sufficient for the chromatic corrections of astrometry and it 
will give photometry of narrow double stars and crowded areas which cannot be obtained from the ground nor can it be obtained with Gaia because the 
long prism photometer spectra of the two stars overlap. Ground-based surveys of multi-colour photometry and spectra will be available for astrophysical 
studies for a large fraction of the stars but Gaia-like precision and accuracy, in the mmag realm, is much better than any ground-based 
survey. There are several reasons to perform photometry:
\begin{itemize}
	\item chromatic corrections of the astrometric observations;
	\item astrophysical information for all objects, including astrophysical 
	classification for objects such as stars, quasars, etc.;
	\item astrophysical characterisation, e.g., interstellar reddening, effective 
	temperatures of stars, photometric red-shifts of quasars, etc.;
	\item photometric distances for those sources with large parallax relative error;
	\item multiwavelength variability of sources.
\end{itemize}

For a new NIR astrometry mission the first reason is still valid to the same extent as for Gaia, but the justification for the others has changed.
By the time of a new NIR astrometry mission, multi-colour photometry will be available for many of the observed stars with better 
spectral resolution than the mission itself can provide. Such photometry will be available from large surveys 
such as Pan-STARRS and LSST, providing five or six spectral bands from 300 to 1100 nm. However, both surveys have bright saturation limits 
at 15 and 17, respectively, and their angular resolution cannot be much better than 0.5 arcsec since they are ground 
based. In addition, they do not cover the full sky. Instead, the new mission will be an all-sky survey and very importantly will be at the 
mmag accuracy level. For Gaia the angular resolution along scan of the astrometric observation is about 0.12 arcsec.
%(FWHM of the sampled line-spread function). 
For photometry the spectral dispersion limits the angular resolution for
multiple stars and crowded regions as the spectra overlap, each spectrum having a length of 1--2 arcsec. 

For the new mission, it is proposed to consider filter photometry which gives higher angular resolution in crowded fields 
due to far less blending and reduces the readout noise and the background contributions, so improving the accuracy at the faint limit. 
The two prisms on Gaia could be replaced by four filters as illustrated in Fig. \ref{fig:bands}.
The first band must cover the complete wavelength (white light) to ensure all stars are detected while subsequent bands have a
reduced wavelength coverage to
provide photometry and calibration data as mentioned above. The filters themselves can be deposited directly onto the detector material
to avoid complicating the telescope optics and there is no need for Sky Mapper (SM) detectors as the motion of the stars on the detectors 
is different for each Field-of-View (FoV) and this can be used to determine their origin and to track them across the focal plane.
This new design was outlined in \href{http://sci.esa.int/future-missions-department/60028-cdf-study-report-gaianir/}{the ESA study report} 
and helped to greatly reduce the complexity and hence cost of the focal plane assembly

Detailed studies of the optimal filter bands need to be done for a mission proposal and the concept in Fig. \ref{fig:bands} may be 
modified depending on a detailed trade-off between achieving accurate chromatic corrections and the possibility of 
getting excellent astrophysical information at the same time.

\subsection{HR diagrams and completeness}\label{sec:photometry:hr}
The multi-band photometry from a new NIR astrometry mission, when combined with its precise parallax, enables the construction of various colour-magnitude diagrams (CMD) from these bands. The less extinction-limited CMDs from GaiaNIR are highly complementary to spectroscopic surveys in two ways. a) The star formation history of any stellar population can be well constrained by the number distribution of stars in the CMDs and better constraining the SFH is the key to better interpreting the chemical evolution of the Milky Way. b) A precise CMD will put a strong prior on log g and effective temperature of the stars and, subsequently, improve the determination of elemental abundances of stars from stellar spectra.

Galactic archaeology is propelled by the myriad of large-scale surveys. However, how well we can disentangle and measure subtle Galactic properties depends critically on our ability to model very precisely the selection function of these surveys -- what stars or populations do we miss, at what location, and at what ratio. Currently, most statistical modelling of the Milky Way relies on the 2MASS IR survey to set the completeness, but the 2MASS completeness cuts off around 16 magnitude. As future spectroscopic surveys such as MSE \citep{2019arXiv190404907T} intend to go deeper than that, having an all-sky, less extinction-limited, deep, NIR survey such as GaiaNIR will be instrumental for statistically modelling the Milky Way.

\subsection{Photometric distances}\label{sec:photometry:photodistances}
Distances of stars are needed not only to derive their 3D location   but to derive tangential velocities from proper motions as well. 
Distances may be obtained from the Gaia parallaxes, but they must be 
supplemented by photometric-spectroscopic distances for remote stars \citep[etc.]{10.1093/mnras/sty990, 2018IAUS..334..153A, 2019arXiv190411302A, 2019MNRAS.484..294D}. The latter 
method will be much strengthened through calibration with accurate trigonometric 
distances from Gaia which will provide, e.g., better than 1.0 percent accuracy for 
10 million stars, most of these will be dwarfs. For giants of luminosity class III, 
the sample within 2 kpc from the Sun will contain stars brighter than V=11.5 if we assume 
M=0.0 mag to be typical. With $\sigma_\pi$=7.1 $\mu$as a relative accuracy better than 7.1/500
or 1.4 percent is then obtained for distances of giants. More precise estimates of the number 
of stars may be obtained from the population synthesis Galaxy star count model TRILEGAL 
\citep{2005A&A...436..895G,2012ASSP...26..165G} as described in Sect.3.1. of \cite{2014ApJ...797...14P}. 

It will be crucial for the determination of photometric distances to these stars that their 
absolute magnitudes can be accurately calibrated. The following studies of photometric calibration 
in clusters are available in \cite{2002ASPC..274..288B}, \cite{2004MNRAS.351.1204V,2004MNRAS.354..815V} and \cite{2007ApJ...661..815R}. 
An accuracy of 10\% will perhaps be common by 2025 after calibration with Gaia data and an accuracy of 
1 or 2 \% may even be obtained for giants and some other types of stars.

The uncertainties of photometry in Gaia DR2 are presented by \cite[][in Figs. 9--11]{2018A&A...616A...4E} and scaling to the end-of-mission,
the photometric errors 
at G=19 are about 3 mmag for G and 10 mmag for the low-dispersion red photometer ($G_\mathrm{RP}$), but much larger for the blue band 
photometer ($G_\mathrm{BP}$).
Absolute magnitudes may therefore be obtained from the spectra of faint red stars \cite{2014arXiv1408.2190H}.
The main source for accurate absolute magnitudes of faint Gaia stars, after luminosity calibration with the Gaia parallaxes, 
would be the many photometric and spectroscopic surveys. Examples include 
4MOST \citep{2012SPIE.8446E..0TD}, 
WEAVE \citep{2014SPIE.9147E..0LD}, 
MOONS \citep{2014SPIE.9147E..0NC}, 
APOGEE I--II \citep{2015arXiv150905420M}, 
LAMOST \citep{2012arXiv1206.3569Z},
etc.

\subsection{Inferred chemical abundances}\label{sec:photometry:chemistry}
Almost every object will have its position on the Kiel and HR diagrams determined with superb accuracy. However, chemistry will be missing, except for a relatively small fraction of objects which will be targeted by ground-based spectroscopic surveys. But knowledge of effective temperature, luminosity and surface gravity allows to play a classical game for estimation of chemical abundances of [Fe/H], [$\alpha$/Fe], and beyond: by using either narrow-band photometric filters centred on strong NIR lines of individual elements (and on the continuum) or using very low resolution spectrophotometry, as done with Gaia's $G_{\rm BP}$ and $G_{\rm RP}$ bands. Studies of this concept are currently ongoing for ESA \citep{2019CoSka..49..320Z} in the context of the Gaia mission where it is being tested on real stars (using Gaia parallaxes, chemistry and parameters derived by the Galah ground-based survey \citep{2018MNRAS.478.4513B}).  Note that ground-based spectroscopic surveys are crucial for providing a training set (comprised of millions of stars) which can then form a base for the billions of sources using suitable artificial intelligence techniques (e.g. The Cannon, \cite{2015ApJ...808...16N}).
If successful this approach can be reformulated for the NIR (e.g. K-band) and is potentially very interesting.
%\\
%
%\noindent
\section{Synergies with other surveys}\label{sec:synergies}

In order to understand how our science case complements other surveys and instruments in the coming decades we briefly review the main relevant ground based surveys currently ongoing and coming into operation. We also consider specific spectroscopic instruments which will provide complementary radial velocity measurements and finally we will consider synergies between the new astrometric mission with some forthcoming space telescope missions currently under preparation. The discussion here is limited to the most obvious projects as new instruments and surveys are being proposed as we write. 

\subsection{Ground based surveys and spectrographs}
\begin{itemize}
	\item SDSS: The Sloan Digital Sky Survey has been one of the most-successful and influential surveys in the history of astronomy, creating three-dimensional maps with deep multi-colour images of one third of the sky, and spectra for more than three million astronomical objects. SDSS-V \citep{2017arXiv171103234K} is unique among spectroscopic surveys in that it is NIR, optical and all-sky, with infrastructures in both hemispheres giving a multi-epoch spectroscopic survey of over six million objects. It is designed to decode the history of the Milky Way, trace the emergence of the chemical elements, reveal the inner workings of stars, and investigate the origin of planets. It will create an integral-field spectroscopic map of the gas in the Galaxy and the Local Group that is a thousand times larger than the current state of the art. SDSS-V's program will start in 2020 and will be well-timed to coincide with other major space missions (e.g., TESS, PLATO, Gaia, eROSITA) and ground-based projects. There are clear synergies between and a new all-sky NIR mission and SDSS-V's Milky Way Mapper (MWM) which is a multi-object spectroscopic survey to obtain NIR and/or visible spectra of more than 4 million stars. 
	
	\item LSST: The Large Synoptic Survey Telescope \citep{2019ApJ...873..111I} is a large, wide-field ground-based system designed to obtain deep images covering the sky visible from northern Chile. The telescope will have an 8.4 m (6.5 m effective) primary mirror, a 9.6 deg$^2$ field of view, a 3.2-gigapixel camera, and six filters (ugrizy) covering the wavelength range 320-1050 nm. Taking repeated images of the sky every few nights in multiple bands for ten years will result in an astronomical catalogue thousands of times larger than previous ones. The project will address: What is the mysterious dark energy that is driving the acceleration of the cosmic expansion? What is dark matter, how is it distributed, and how do its properties affect the formation of stars, galaxies, and larger structures? How did the Milky Way form, and how has its present configuration been modified by mergers with smaller bodies over cosmic time? What is the nature of the outer regions of the solar system? Is it possible to make a complete inventory of smaller bodies in the solar system, especially potentially hazardous asteroids that could someday impact the Earth? Are there new exotic and explosive phenomena in the universe that have not yet been discovered? A new NIR mission would broaden the wavelength range and provide much more accurate astrometry down to H~$<20$ magnitude. LSST will do good astrometry but with limited accuracy from the ground, obtaining parallax and proper-motion measurements of comparable accuracy, $\sigma_{\rm \pi}\sim0.6$~mas, $\sigma_{\rm \mu}\sim0.2$~mas~yr$^{-1}$, to those of Gaia at its faint limit (G<20) and smoothly extend the error versus magnitude curve deeper by about 5 mag \citep{2013AAS...22124706I}. Thus, the LSST astrometric project is anchored in the successful Gaia mission and is extremely complementary to GaiaNIR.
	
	\item Pan-STARRS is a wide-field imaging system operated by the Institute for Astronomy at the University of Hawaii. The survey uses a 1.8 meter telescope and its 1.4 gigapixel camera to image the sky in five broadband  filters (g, r, i, z, y) sensitive from 400--1100 nm \citep{2012ApJ...750...99T}. By detecting differences from previous observations of the same areas, Pan-STARRS is discovering a large number of new asteroids, comets, variable stars, supernovae and other celestial objects. Its primary mission is now to detect Near-Earth Objects that threaten impact events and it is expected to create a database of all objects visible from Hawaii (three quarters of the entire sky) down to apparent magnitude 24. Similar synergies with a new NIR mission apply here.
	
	\item VVV: ESO's NIR variability survey \citep{2010NewA...15..433M} is scanning the Milky Way bulge and an adjacent section of the mid-plane where star formation activity is high. The survey uses the 4-metre VISTA telescope during five years, covering $\sim$~10$^9$ point sources across an area of 520 deg$^2$, including 33 known globular clusters and $~\sim$~350 open clusters. The final product is a deep NIR atlas in five passbands (900-2500 nm) and a catalogue of more than 10$^6$ variable point sources. Unlike single-epoch surveys that, in most cases, only produce 2-D maps, the VVV variable star survey will enable the construction of a 3-D map of the surveyed region using distance indicators (albeit with potentially unknown biases \citep{2018Ap&SS.363..127M}) such as RR-Lyrae stars, and Cepheids, yielding important information on the ages of the populations. An extended survey (VVVX) is ongoing. Observation data will be combined from many sources including Gaia to better understand variable sources in the inner Milky Way, globular cluster evolution, a population census of the Galactic Bulge and centre, as well as studies of star forming regions in the disk. By combining VVV, whose accuracy is currently around $0.7$~mas~yr$^{-1}$ \citep{Smith2018}, and the precise distances from a new NIR astrometry mission one would get more precision in extinction and distances in the most difficult and unexplored regions of the Milky Way disk which would also help to calibrate any biases in the VVV distance indicators.

	%\citep[,][]{2012SPIE.8446E..0TD}, on VISTA   390 -  100nm \\
	%\item 4MOST gives spectra for radial velocities plus photometry for classification and distances of faint stars may be obtained in great number 
	\item  4MOST is the ESOs wide-field, high-multiplex (1500--3000 fibres) spectrograph \citep[see Fig. 1][]{2012SPIE.8446E..0TD} 
	which will be running nearly full time on the 4m-class telescope (VISTA) from about 2022, and capable of providing radial velocity, stellar parameters and chemical composition for a large number of faint stars. Radial velocities accuracies of $\le$~2~km~s$^{-1}$ for the 
	faintest stars observed by Gaia, which will be also observed in a new NIR astrometry mission, may be obtained with 4MOST, which will provide
	the full 3-D space velocities for unambiguous dynamical studies as well as chemistry for more than 10 million stars \citep{2019Msngr.175...30C, 2019Msngr.175...35B}.
	
	%\citep[WEAVE,][]{2014SPIE.9147E..0LD}, on William Herschel Telescope  370 - 1000nm \\
	\item WEAVE \citep{2014SPIE.9147E..0LD} is a new wide-field multi-object spectrograph for the William Herschel Telescope and will 
	do optical ground-based follow up of the LOFAR and Gaia surveys. WEAVE will host 1000 multi-object fibres which are fed to a 
	single spectrograph, with a pair of cameras. The project is now being developed with 
	commissioning expected in 2020. Unfortunately, 4MOST and WEAVE will only operate in the wavelength 
	region 390--1000~nm and 370--1000~nm respectively and will not probe the NIR to open new science cases in the Galactic disk and the Galactic centre region.
	
	%\citep[MOONS,][]{2014SPIE.9147E..0NC}, on VLT  800-1800nm \\
	\item MOONS: Multi-Object Optical and NIR Spectrograph \citep{2014SPIE.9147E..0NC} on the Very Large Telescope (VLT) 8.2 meter 
	telescope will operate in the wavelength range 800--1800~nm. The large collecting area offered by the VLT, combined with the 
	large multiplex and wavelength coverage will constitute a powerful, unique instrument able to pioneer a wide range of Galactic, extragalactic 
	and cosmological studies and provide follow-up for major facilities such as Gaia. MOONS is capable of providing radial velocity and chemical compositions 
	for faint stars for a new NIR astrometry mission.
	
	%\citep[APOGEE I-II,][]{2015arXiv150905420M}, 1510-1700 nm \\
	\item APOGEE--I \& II \citep{2015arXiv150905420M}, part of the SDSS survey, operates from 1510--1700~nm using high-resolution, high signal-to-noise IR spectroscopy to	penetrate the dust that obscures the disk and bulge of our Galaxy to survey over 400,000 red giant stars across the full range of the Galactic bulge, bar, disk, and halo. By the end of the survey in 2020 APOGEE will provide a catalogue of 600,000 objects in both hemispheres. 
	Combining APOGEE data with Gaia is already having a major impact but this is just the tip of the iceberg as the statistics are low both because the number of targets are still low, and because Gaia is not NIR. APOGEE will yield $\sim$0.1 km/s accuracy radial velocities as well as chemical information for over 15 elements.
	
	\item Subaru Prime Focus Spectrograph (PFS) \citep{2014PASJ...66R...1T} is a massively multiplexed fiber-fed optical and NIR three-arm spectrograph (Nfiber = 2400, $380\leq \lambda \leq 1260$ nm 1.3 degree diameter field of view). Data will be secured for $10^6$ stars in the Galactic thick-disk, halo, and tidal streams as faint as V $\sim$ 22, including stars with V < 20 to complement the goals of the Gaia mission. A medium-resolution mode with R = 5000 to be implemented in the red arm will allow the measurement of multiple $\alpha$-element abundances and more precise velocities for Galactic stars.
	
	\item The Large Sky Area Multi-Object Fiber Spectroscopic Telescope (LAMOST) \citep{2012arXiv1206.3569Z} survey contains two main parts: the LAMOST ExtraGAlactic Survey (LEGAS), and the LAMOST Experiment for Galactic Understanding and Exploration (LEGUE) survey of Milky Way stellar structure. The unique design of LAMOST enables it to take 4000 spectra in a single exposure to a limiting magnitude as faint as r=19 at the resolution R=1800. This telescope therefore has great potential to efficiently survey a large volume of space for stars and galaxies.
	
	\item And other surveys are coming online, a few examples include: The Maunakea Spectroscopic Explorer (MSE) \citep{2019arXiv190404907T}
    is an end-to-end science platform for the design, execution and scientific exploitation of spectroscopic surveys; SpecTel
    \citep{2019arXiv190706797E} is a proposed $10$--$12$ meter class Spectroscopic Survey Telescope which is being designed to access a larger fraction of objects from Gaia, LSST, Euclid, and WFIRST than any currently funded or planned spectroscopic facility.  
	%The Dark Energy Spectroscopic Instrument (DESI) \citep{2019arXiv190101581V} will measure the effect of dark energy on the expansion of the universe. It will obtain optical spectra for tens of millions of galaxies and quasars.
\end{itemize}

\subsection{Space Missions}
\begin{itemize}
	\item WFIRST: The Wide Field Infrared Survey Telescope \citep{2015arXiv150303757S} is a NASA observatory which will perform galaxy surveys over thousands of square degrees for dark energy weak lensing and baryon acoustic oscillation measurements and will monitor a few square degrees for dark energy SN Ia studies. It will perform microlensing observations of the galactic bulge for an exoplanet census and direct imaging observations of nearby exoplanets with a pathfinder coronagraph. WFIRST has a 2.4~m telescope, the same size as Hubble’s, but with a FoV 100 times greater. WFIRST is slated for launch in the mid-2020s although it faces repeated funding challenges. The wide field instrument includes filters that provide an imaging mode covering 480--2000 nm, and two slit-less spectroscopy modes covering 1000--1930 nm with resolving power 450--850, and 800--1800 nm with resolving power of 70--140. The instrument provides a sharp point spread function, precision photometry, and stable observations for implementing the dark energy, exoplanet microlensing, and NIR surveys. WFIRST is not an all-sky survey mission but is extremely complementary to Gaia and the JWST and could act as a source of first epoch measurements for a new NIR astrometry mission in a similar manner to small-JASMINE (see below).
	
	\item The Euclid mission \citep{2013LRR....16....6A} will have very high quality NIR photometry but its science cases are focused on the extragalactic sky which is presently defined by the regions covering $\left|b\right|>30^\circ$ which precisely avoids those regions which would be most interesting for NIR Galactic astronomy. Euclid measurements of objects out of the Galactic plane could act as second epoch measurements for Gaia and for very red faint objects, not already in Gaia, could provide first epoch measurements for a new NIR astrometry mission.
	
	\item JWST: The James Webb Space Telescope \citep{2005ASPC..338...59N} has a NIR (600--2500 nm) instrument called NIRCam with 32 mas pixels. Combined with the telescope's 6.5 meter segmented primary mirror, NIRCam images will have an angular resolution similar to what is achieved by the HST. Thus, JWST will have the potential for astrometric performance similar to that achieved with HST imaging. Using archived data from HST observations to provide first epoch positions, JWST will measure proper motions of stars within globular clusters and other crowded fields. JWST can also provide parallaxes for nearby objects too faint (neutron stars, L-dwarfs, e.g.) for Gaia. As an example of the usefulness of all-sky astrometry the Gaia astrometric catalog is being included in the HST Guide Star Catalog (GSC), the Hubble Source Catalog (HSC) and the JWST calibration field catalog in order to improve their astrometric accuracy \citep{sahlmann_nelan_chayer_mclean_lallo_2017}, similar applications of a new astrometry mission will surely arise. As with WFIRST the JWST can be used to provide first epoch measurements in small regions of the sky. However, a new all-sky NIR mission would provide astrometric measurements with slightly lower resolution all over the sky!
	
	\item LISA: The Laser Interferometer Space Antenna \citep{2013GWN.....6....4A} is a mission led by ESA to detect and accurately measure gravitational waves from astronomical sources. LISA would be the first dedicated space-based gravitational wave detector. Ultra-short period Galactic double white dwarf binaries are unique multi-messenger tracers of the Milky Way. They can be detected in large numbers through electromagnetic radiation by Gaia and through gravitational waves by the upcoming LISA mission \citep{valeriya_korol_2019_3237213}. The synergies between Gaia and LISA observations of double white dwarf binaries would allow the study of the Milky Way baryonic structure due to LISA's ability to localise binaries through virtually the whole Galactic plane while Gaia yields information on their motion; tracing the underlying total enclosed mass. However, many of the objects detected by LISA will be in regions that are hardly accessible to visible observations such as the inner part of the Galactic disc, the bulge, and beyond \citep{10.1093/mnras/sty3440}. Of course such synergies can be improved if a new NIR astrometry mission can partially see into the disk and centre of the Galaxy where many of the double white dwarfs are found.
	
	\item PLATO \citep{2014ExA....38..249R, 2018SPIE10698E..4XM, 2017AN....338..644M}, TESS \citep{2014SPIE.9143E..20R} and beyond: There are
    also synergies with the many upcoming exoplanet surveys, generally space astrometry addresses very long period massive exoplanets, while
    other transit and radial velocity exoplanet surveys are aimed at shorter period exoplanets. We have shown using numerical studies (see
    section \ref{sec:nir:exoplanets}) that the combination of two missions with a 20 year gap can yield exoplanets with periods of
    30--40 years allowing the co-planarity of these systems to be studied with multi-mission data. Therefore, a new astrometry mission will be very complementary for exoplanet science particularly as astrometry recovers all seven orbital parameters for the exoplanet as well as for the science being made with asteroseismology for Galactic archaeology \citep{2017AN....338..644M}.
	
	\item Small-JASMINE \citep{2013IAUS..289..433Y} is being developed by National Astronomical Observatory of Japan, Japanese universities and ISAS/JAXA in collaboration with ESA, Gaia-DPAC and US Naval Observatory. The aim is to make astrometric measurements in the Hw-band (1100--1700 nm). The mission will use the Gaia visible reference frame to do relative plate overlap astrometry in a radius of 0.7 degrees or $\sim$~100 pc around the bulge and Galactic centre and selected directions towards interesting objects such as star forming regions and/or M-dwarf exoplanet transit follow-up and make high frequency measurements (100 minutes) with astrometry accuracy of $~\sim$~25~$\mu$as (yr$^{-1}$) for bright stars (Hw$<12.5$~mag) and $\sim125$~mas~yr$^{-1}$ for fainter stars (Hw$<15$~mag). Small-JASMINE will be very complementary to Gaia providing proper motions for about hundred thousand bulge stars allowing the astrophysics of the Galactic centre to be studied. The mission is due to be launched by mid-2020 and can be seen as a precursor by providing first epoch measurements, albeit very limited, for an all-sky visible and NIR survey. The combination of Small-JASMINE--GaiaNIR in the Galactic nuclear bulge region would be a very powerful tool.
\end{itemize}

The combination of multi-messenger data from these surveys and instruments would greatly complement a new space astrometry mission in the
NIR in many cases providing first epoch measurements.  The spectroscopic instruments would provide the third dimension of the space
velocities needed particularly in the dusty regions that currently cannot be explored by Gaia and at first sight it would seem unnecessary
to include a spectrograph in a new mission design. However, a new NIR astrometry mission could provide spectra for possibly billions of objects (to some
limiting magnitude) in the visible and NIR globally on the sky, albeit at low resolution, which would provide this third component of the
space velocity for many observed objects compared to some tens of millions of spectra from selected areas with different instruments. A new
mission would potentially provide a unique opportunity to obtain photometry and spectroscopy with a single instrument making the calibration
of the science results uniform across the sky. However, we have concluded that the spectroscopic option for a new mission is a nice addition
(presumably at considerable cost) but is not a necessary requirement to achieve the science goals outlined in this document.

In this context it is worth noting that the advent of Gaia led to the community organizing facilities for Gaia follow-up (see the
\href{http://www.stecf.org/coordination/eso-esa/galpops.php}{ESA-ESO WG document}) from where projects like 4MOST, WEAVE, and the Milky Way
Mapper project in SDSS-V were born.  We can thus expect the astronomy community to undertake similar efforts to organize a massive NIR
spectroscopic follow-up should GaiaNIR become a reality.

\section{Mission scenarios}\label{sec:mission} %{\color{red} Max 1 page -- being prepared by Hobbs.}

As discussed in the introduction we see international collaboration as the best way to achieve the science goals outlined here while keeping within the M-class budget
constraints of ESA, so collaboration with the US, Australia and/or Japan is an essential ingredient. The US is the world leader in 
detector technology particularly for the NIR so a collaboration with the US makes a lot of sense. We have recently submitted an APC white paper \citep{2019arXiv190705191H} to the ASTRO-2020 call outlining state of the profession considerations on detector development for our application, which may also be generally applicable 
to astronomy, remote sensing, planetary observation and for LIDAR applications. In the APC white paper four different technologies have been identified which could meet our needs
provided sufficient market motivation and investment in them can be found. Here we make a clear statement that it is possible to have visible-NIR detectors 
with TDI like operation but their development will come with a financial cost.

\begin{enumerate}
	\item A hybrid solution which uses a HgCdTe NIR detector layer bump bonded to a Si CCD. The idea is that the photons are detected in the surface NIR layer and transferred to the Si buried channel at each pixel. Charge can then be easily moved along the pixels in sync with the charge generation, thus achieving TDI. What is not known yet is how efficiently the charge can be transferred from the NIR detection layer to the Si CCD and if both materials can be operated at the same temperature in a space environment making this development potentially complex.
	\item Using HgCdTe Avalanche Photodiodes (APDs) with TDI-like signal processing capability. The challenge here is to scale the existing technology to larger format arrays and ensure the dark current does not introduce unwanted noise at temperatures above 100 K. The Australia National University working with Leonardo (Italy) are actively developing this technology for large-format astronomy applications and are keen to be involved. The development effort required for these devices does not seem excessive.
	\item Ge detectors due to the lower band gap can detect NIR radiation of longer wavelengths than possible with Si detectors. Clearly this technology is new but many of the manufacturing techniques developed for Si are also applicable to Ge and further development is needed to see if they can be used for low noise and visible-NIR capabilities in large format arrays. The wavelength range is, however, limited to 1600~nm.
	\item Microwave Kinetic Inductance Detectors (MKIDs) are cooled, multispectral, single photon counting, multiplexed devices capable of observation in the UV 
	through to mid-wave IR. They measure the energy of each photon to within several percent and log the time of arrival to within 2 microseconds, making them ideal for TDI like operation. 
	Whilst relatively new, small MKID arrays (several thousand pixels) have already been utilised on ground-based telescopes but moving towards Gigapixel space based devices is very challenging given that they also require active cooling which is undesirable.
\end{enumerate}
The use of visible-NIR TDI capable detectors is optimal for maximising the science return by reusing the same well established concepts as
Gaia. A new NIR astrometry mission will detect at least 5 times as many stars and open up the important NIR regions of the Galaxy. While the spacecraft is
scanning outside the Galactic plane it would be possible to go deeper than $G=20.7$ improving the science return in the Halo region. The main challenge for this exciting science mission is clearly the development of the new detector technology at a reasonable cost.

\newpage
\appendix
\section*{Members of the core proposing team}\label{sec:proposers}
	
\begin{longtable}[l]{ll}
	David Hobbs                \ignore{& david@astro.lu.se                     }& Lund Observatory, Lund University, Box 43, SE-22100, Lund, Sweden\\
	Anthony Brown              \ignore{& brown@strw.leienuniv.nl               }& Leiden Observatory, Leiden University, Niels Bohrweg 2, 2333 CA Leiden, The Netherlands\\
	Erik H{\o}g                \ignore{& ehoeg@hotmail.dk                      }& Niels Bohr Institute, University of Copenhagen, Blegdamsvej 17, 2100 Copenhagen, Denmark \\
	Carme Jordi                \ignore{& carme@fqa.ub.edu                      }& Institut de Ci\`encies del Cosmos, ICCUB-IEEC, Spain \\
	Daisuke Kawata             \ignore{& d.kawata@ucl.ac.uk                    }& Mullard Space Science Laboratory, University College London, Holmbury St. Mary, Dorking RH5 6NT, UK\\
	Paolo Tanga                \ignore{& Paolo.Tanga@oca.eu                    }& Observatoire de la Côte d'Azur, 06304 Nice Cedex 4, France\\
	Sergei Klioner             \ignore{& Sergei.Klioner@tu-dresden.de          }& Lohrmann Observatory, Technical University Dresden, Germany \\
	Alessandro Sozzetti        \ignore{& alessandro.sozzetti@inaf.it           }& INAF-Osservatorio Astrofisico di Torino, Via Osservatorio 20, 10025 Pino Torinese, Italy\\
    {\L}ukasz Wyrzykowski      \ignore{& lw@astrouw.edu.pl                     }& Warsaw University Observatory, Al. Ujazdowskie 4, 00-478 Warszawa, Poland\\
	Nic Walton                 \ignore{& naw@ast.cam.ac.uk                     }& Institute of Astronomy, University of Cambridge, Madingley Road, Cambridge, CB3 0HA, UK\\
	Antonella Vallenari        \ignore{& antonella.vallenari@inaf.it           }& Osservatorio Astronomico di Padova, INAF, Vicolo dell’Osservatorio 5, I-35122 Padova, Italy \\
	Valeri Makarov             \ignore{& valeri.makarov@gmail.com              }& U.S. Naval Observatory, 3450 Massachusetts Avenue NW, Washington, DC 20392, USA\\
	Jan Rybizki                \ignore{& rybizki@mpia.de                       }& Max Planck Institute for Astronomy, K\"onigstuhl 17, D-69117 Heidelberg, Germany\\
	Fran Jim\'enez-Esteban     \ignore{& fran.jimenez-esteban@cab.inta-csic.es }& Centro de Astrobiolog\'{i}a (CSIC-INTA), Camino Bajo del Castillo s/n, 28692 Villanueva de la Ca\~nada, Spain \\
	Jos\'e A. Caballero        \ignore{& caballero@cab.inta-csic.es            }& Centro de Astrobiolog\'{i}a (CSIC-INTA), Camino Bajo del Castillo s/n, 28692 Villanueva de la Ca\~nada, Spain \\
	Paul J. McMillan           \ignore{& paul@astro.lu.se                      }& Lund Observatory, Lund University, Box 43, SE-22100, Lund, Sweden\\
	Nathan Secrest             \ignore{& nathansecrest@msn.com                 }& U.S. Naval Observatory, 3450 Massachusetts Avenue NW, Washington, DC 20392, USA \\
	Roger Mor                  \ignore{& rmor@fqa.ub.edu                       }& Institut de Ci\`encies del Cosmos, ICCUB-IEEC, Spain \\
	Jeff J. Andrews            \ignore{& jeff.andrews@nbi.ku.dk                }& Niels Bohr Institute, University of Copenhagen, Blegdamsvej 17, 2100 Copenhagen, Denmark \\
	Toma\v z Zwitter           \ignore{& tomaz.zwitter@fmf.uni-lj.si           }& University of Ljubljana, Faculty of Mathematics and Physics, Ljubljana, Slovenia \\
	Cristina Chiappini         \ignore{& cristina.chiappini@aip.de             }& Leibniz-Institut für Astrophysik Potsdam (AIP), An der Sternwarte 16, 14482 Potsdam, Germany \\
	Johan P. U. Fynbo          \ignore{& jfynbo@nbi.ku.dk	                   }& Niels Bohr Institute, University of Copenhagen, Blegdamsvej 17, 2100 Copenhagen, Denmark \\
	Yuan-Sen Ting              \ignore{& ting.yuansen.astro@gmail.com          }& Institute for Advanced Study, Princeton, NJ 08540, USA \\
	Daniel Hestroffer          \ignore{& Daniel.Hestroffer@obspm.fr            }& IMCCE  -  Observatoire de Paris, F-75014 PARIS, France \\	
	Lennart Lindegren          \ignore{& lennart@astro.lu.se                   }& Lund Observatory, Lund University, Box 43, SE-22100, Lund, Sweden\\
	Barbara McArthur           \ignore{& mca@astro.as.utexas.edu               }& McDonald Observatory, University of Texas at Austin, Austin, TX 78712-1205, USA\\
	Naoteru Gouda              \ignore{& naoteru.gouda@nao.ac.jp               }& National Astronomical Observatory of Japan, 2-21-1, Osawa, Mitaka, Tokyo 181-8588, Japan\\
	Anna Moore                 \ignore{& Anna.Moore@anu.edu.au                 }& The Australian National University, Canberra, ACT 2611, Australia\\
	Oscar A. Gonzalez          \ignore{& oscar.gonzalez@stfc.ac.uk             }& STFC UK Astronomy Technology Centre, The Royal Observatory Edinburgh, EH9 3HJ, UK \\
	Mattia Vaccari             \ignore{& mattia.vaccari@gmail.com              }& University of the Western Cape, Robert Sobukwe Road, 7535 Bellville, Cape Town, South Africa \\

 	\\
\end{longtable}

\newpage
\bibliographystyle{aa}
\bibliography{Voyage2050WhitePaper}

\end{document}